\documentclass[12pt]{article}

\usepackage[utf8]{inputenc}
\usepackage[T1]{fontenc}
\usepackage{lmodern}
\usepackage{graphicx}
\usepackage[figurename=Fig.,labelfont=bf,labelsep=period]{caption}
\usepackage{subfig}
\usepackage{caption}
\usepackage{amsmath}
\usepackage{newtxtext,newtxmath}
\usepackage[colorlinks=true,citecolor=red,linkcolor=black]{hyperref}
\usepackage{booktabs}
\usepackage{enumerate}
\usepackage{apacite} 
\usepackage{hyperref} 
\usepackage{tikz}
\usetikzlibrary{shapes.geometric,arrows,calc,fit,backgrounds,shapes.multipart}
\usepackage{placeins}
\usepackage{authblk}

\title{\vspace{-15mm} Predicting high rate granular transition and fragment statistics at the onset of granular flow for brittle ceramics}
\date{}
\author[1]{\large Amartya Bhattacharjee\thanks{abhatt15@jhu.edu}}
\author[2]{\large Ryan C. Hurley\thanks{rhurley6@jhu.edu}}
\author[1]{\large Lori Graham-Brady\thanks{lori@jhu.edu}}
\affil[1]{\footnotesize Department of Civil and Systems Engineering, Johns Hopkins University, Baltimore, Maryland 21218, USA}
\affil[2]{\footnotesize Department of Mechanical Engineering, Johns Hopkins University, Baltimore, Maryland 21218, USA}
\begin{document}

\maketitle
\begin{abstract}
Brittle materials under impact loading exhibit a transition from a cracked solid to a granular medium. Appropriate representation of this transition to granular mechanics and the resulting initial fragment size and shape distribution in computational models is not well understood. The current work provides a numerical model to analyze competitive crack coalescence in the transition regime and provides insight into the onset of comminution and the initial conditions for subsequent granular flow. Crack statistics obtained from initial flaws using a wing crack growth based damage model have been used to discretely model elliptical cracks in three dimensions, with and without a minimal intersection constraint. These cracks are then allowed to coalesce with nearby cracks along favourable directions and the output fragment statistics have been predicted. The evolving fragmentation offers insight into the onset of comminution as well as the final transition to granular mechanics and the resulting initial fragment statistics. A simple phenomenological model has been proposed that suggests a transition criterion resembling the one obtained from the numerical model.
\end{abstract}

\section{Introduction}
In ceramic armour systems subjected to penetration, rock blasting or asteroid impacts, typical strain rates are much higher than in classical static wing crack growth models \shortcite{ASHBY1986}. In such applications, inertial effects dominate the crack growth, and the dynamic stress intensity factor \shortcite{NEMATNASSER1994} is used as a measure to determine the rate of crack growth. While for static cases in uniaxial compression, only the most favourable cracks grow into macro-cracks and cause failure \shortcite{ASHBY1986}, dynamic crack growth causes simultaneous growth of a population of cracks. This renders weakest link failure models \shortcite{JiaLiang2012} based on macro-crack initiation unsuitable for dynamic compressive loading. Modelling dynamic crack growth either involves tracking the growth of discrete cracks \shortcite{Falk2001,Daux2000,ABEDI2017}, crack bands \shortcite{Bazant1983,JiaLiang2016}, phase field models \shortcite{SPATSCHEK2011,HOFACKER2012,BORDEN2012,SCHLUTER2014}, or continuum damage models that track the growth of crack populations with homogenization schemes to account for the evolving damaged material properties \shortcite{PALIWAL2008,KATCOFF2014}.  Phase field models can be used to accurately predict the three dimensional micro-cracking patterns in quasi-static brittle failure \shortcite{NGUYEN2016} and can be extended to account for the various failure modes and mechanisms in dynamic brittle fragmentation. However, armor ceramics have very high initial defect densities, in the form of pores and inclusions, which serve as initiation sites for cracks. Modelling the interaction and simultaneous propagation of millions of cracks, as is the case for high rate loading of armor ceramics, can become prohibitively expensive. This is when continuum damage models are particularly useful over discrete modeling of cracks, crack bands or phase field models. However, the applicability of such continuum-based models in the granular transition regime, in which the cracks approach the element size, is questionable.\par
One viewpoint on the growth of cracks in the granular transition regime assumes that, as these cracks grow, they interact with one another, influencing their growth, which manifests not only through changes in effective properties of the surrounding matrix \shortcite{NEMATNASSER1994,Grechka2006} but also in changes to the local stress field and the direction of crack growth. When such changes occur, one is dealing with simultaneous growth and coalescence of multiple cracks. These coalesced cracks now behave as larger individual cracks. Eventually many of these cracks connect with one another to form a network of connected cracks that fragment the domain into smaller particles. This process involves a competition not only between isolated wing crack growth and crack coalescence but also between the different modes of crack coalescence \shortcite{WONG1998,Taoying2017} and the growth of secondary cracks \shortcite{BOBET2000}.\par
An alternate viewpoint of crack growth in the granular transition regime is that fragmentation is driven by crack branching \shortcite{Astrom1997,Kekalainen2007}. Crack branching requires higher crack velocities ($0.43-0.46C_R$ \shortciteA{Katzav2007}). For high crack densities, crack coalescence might occur before such crack velocity is reached and any significant crack branching might have occurred. Under such situations, crack branching alone might not influence the fragmentation process. Right before the onset of fragmentation, the length scale of cracks can be expected to be comparable to the spacing, making homogenization of damaged matrix properties based on dilute approximation questionable. \par
The fragment size distribution resulting from the fragmentation process has been observed over two different length scales \shortcite{HOGAN2016,HOGAN2017}: a) the length scale of the defects  (Regime I), b) the macroscopic scale (Regime II). Scanning electron microscope images of the fracture surface, for Regime I fragments, showed defects located on the fracture surface and not inside the material. This suggests transgranular fracture with the microstructural defects serving as crack initiation sites. On the contrary, Regime II fragments are a consequence of transverse and axial splitting macro-cracks, which are influenced by the specimen geometry, boundary conditions and the loading rate. For a projectile impacting a ceramic plate, the Mescall zone at the tip of the penetrator or the slip zone in earthquake faults \shortcite{BenZion2008} is composed of heavily comminuted material that undergoes granular flow. This is analogous to the Regime I fragmentation observed in \shortcite{HOGAN2016,HOGAN2017}.\par
\shortciteA{CHOCRON2012,KRIMSKY2019} show that the strength as well as the failure mechanism of thermally shocked boron carbide with more initial defects is different from that of the pristine material. This hints towards an obvious dependence of the failure criterion with microstructural defects. This might also mean that calibration of initial conditions for granular flow in the comminuted zone, from initial fragment statistics obtained from thermally shocked samples might not be accurate.\par
Understanding fracture and fragmentation in the post peak strength region of ceramics as they transition on the granular mechanics yield surface from a high strength to low strength regime is a non-trivial task. The instabilities associated with crack growth in a competitive environment and the multiple possible modes of crack coalescence are not well understood and are difficult to address. In most cases, the transition to granular phase has been modelled either through metrics specific to a model or non-physical threshold of physical quantities \shortcite{DESHPANDE2011,TONGE2016}. In \citeauthor{DESHPANDE2011}, the transition from lattice plasticity to granular physics is modelled as a gradual transition described by a damage parameter. When this parameter reaches unity, the material is fully granular. In some models \shortcite{JOHNSON1994}, a critical stress threshold has been used to mark the transition to a completely failed, granular like solid. Typical transitional damage values in models, estimated from crack statistics, might be set at a lower value than what might actually lead to failure or granular transition. Often these are based on calibration experiments and/or signify the limit of applicability of continuum assumption \shortcite{TONGE2016}. Quantifying crack lengths for a network of intersecting cracks is often impossible, and a more accurate damage quantification might be in terms cracked surface area per unit volume \shortcite{KRIMSKY2019}.  \shortciteA{LYAKHOVSKY2011,LYAKHOVSKY2014} has tried to model the granular phase transition using a Continuum Damage Breakage Mechanics (CDBM) model using a critical damage threshold ($\alpha$)  expressed as function of the strain invariant ratio ($\xi=I_1/\sqrt{I_2}$). The study concludes that depending on the loading conditions, a damaged solid can transition to a pseudo-liquid granular flow phase or a pseudo-gas fragmentation phase. The damage threshold ($\alpha$)   in the model is a parameter that represents the state of the material. It varies between $\alpha=0$ for a material without any damage to $\alpha=1$ for a fully damaged material and affects the elastic constants in the model. The model predicts a rapid transition from a fragmented stage to granular flow. \shortciteA{Kun1999} predicts a sharp transition of colliding solids to fragmentation based on the critical threshold of the impact energy to “binding energy” ratio for colliding solids.\par 
It should be noted, however, that fragmentation can occur well before granular transition, and may be influenced by the specimen geometry. In such cases the peak strength is limited by fragmentation and not due to the competition between softening of the modulus and increase in stress due to increase in strain. This can be expected in unconfined Kolsky bar experiments, where structural fragmentation (Regime II) can limit stress buildup. This kind of fragmentation has not been studied in this work.\par
In this work, the transition to granular mechanics for ceramics under high rate loading conditions has been addressed.  Crack distributions are first obtained from a wing crack growth-based continuum model for uniaxial loading conditions. Two different algorithms for three dimensional crack coalescence are then used. The outputs of the algorithms have been used to predict the transition to granular phase as well as the resulting fragment size distribution at the onset of granular flow. The transition criterion has been expressed in terms of an equivalent crack length or a damage-stress combination. A phenomenological transition model has also been proposed, which suggests a similar form for the granular phase transition criterion, and can be used in continuum brittle fragmentation codes, to capture the change from comminuted ceramic to a granular medium, and provide input parameters for subsequent granular flow. Although the algorithms and transition model rely on uniaxial loading conditions, with a proper crack growth criterion and 3D-anisotropic damage model \shortcite{HU2015,KOLARI2017}, they can be extended to multi-axial loading.

\begin{table}[h!]
\begin{center}
\begin{tabular}{l|c|c|r}
\toprule
\textbf{Parameter} & \textbf{Variable} & \textbf{Units} & \textbf{Value}\\
\midrule
Defect Density (random orientation) & $\eta$ & $m^{-3}$ & $22x10^{12}$\\
Defect size & $l_i$ & $\mu$ & 10\\
Fracture Toughness & $K_{IC}$ & $MPa\sqrt{m}$ & 2.5\\
Strain Rate & $\dot{\epsilon}$ & $s^{-1}$ & $10^6$\\
Elastic Modulus & $E$ & $GPa$ & 461\\
Density & $\rho$ & $kg/m^3$ & 2520\\
Poisson's Ratio & $\nu$ & & 0.177\\
Coefficient of friction & $\mu_{flaw}$ & & 0.8\\
\bottomrule
\end{tabular}
\caption{Default material and model parameters (Defect density values correspond to random orientation of cracks)}
\label{tab:table1}
\end{center}
\end{table}

\section{Methodology}
This section details the various steps involved in the fragmentation model, summarized in Fig:\ref{fig:FlowChart}. The model involves the simulation of the cracked microstructure at a given instant using a three dimensional voxelized space called the simulation box. Each crack is represented by a collection of connected voxels. The simulation box contains many such cracks. As the microstructure becomes progressively more cracked, neighboring cracks start coalescing with one another. Gradually the microstructure is transformed into a network of three dimensional connected voxels representing cracked material. A connected region of voxels representing uncracked material, completely enclosed by voxels representing cracked  material is a fragment. Sec: \ref{ssec:CrackSizeStat} discusses evaluating the initial instantaneous statistics of crack populations from initial defects using a wing crack growth based damage model. Three dimensional cracks are then simulated from these crack statistics as explained in Sec: \ref{ssec:CrackSim3D}. Sec: \ref{sssec:CrackGrowth} discusses a 2D crack coalescence problem using stress intensity factor based calculations. Sec:\ref{ssec:CrackCoa} attempts to model crack coalescence due to further crack growth using two different approaches: (a) Coalescence surface approach (Sec:\ref{sssec:CoaSurfApp}), (b) Coalescence zone approach (Sec:\ref{sssec:CoaZoneApp}). Sec: \ref{sssec:ThreshDist} explains the choice of a threshold distance for crack coalescence. The two approaches have been compared in Sec: \ref{sssec:CoaCompApp}. Finally, Sec: \ref{sssec:DilFragStat} discusses a connected region algorithm to extract fragments followed by a dilation procedure on the connected regions to compensate for any loss of material mass due to resolution size.

\tikzstyle{startstop} = [rectangle, rounded corners, minimum width=0.8cm, minimum height=0.8cm,text centered, draw=black, fill=red!20]
\tikzstyle{io} = [trapezium, rounded corners, trapezium left angle=80, trapezium right angle=-80, minimum width=4cm, text centered, draw=black, fill=blue!20]
\tikzstyle{io1} = [rectangle split, rectangle split parts=2, minimum width=0.5\linewidth, minimum height=0.8cm, rectangle split part align={center,justify}, rounded corners, rectangle split part fill={blue!20,gray!15}, rectangle split draw splits=false, draw=black]
\tikzstyle{process1} = [rectangle split, rectangle split parts=2, minimum width=0.8cm, minimum height=0.8cm, rectangle split part align={center,justify}, rounded corners, rectangle split part fill={orange!30,gray!15}, rectangle split draw splits=false, draw=black]
\tikzstyle{process} = [rectangle, rounded corners, minimum width=0.8cm, minimum height=0.8cm,text centered, draw=black, fill=orange!30]
\tikzstyle{decision} = [diamond, rounded corners, aspect=2, minimum width=0.8cm, minimum height=0.8cm, text centered, draw=black, fill=yellow!60]
\tikzstyle{arrow} = [thick,->,>=stealth]
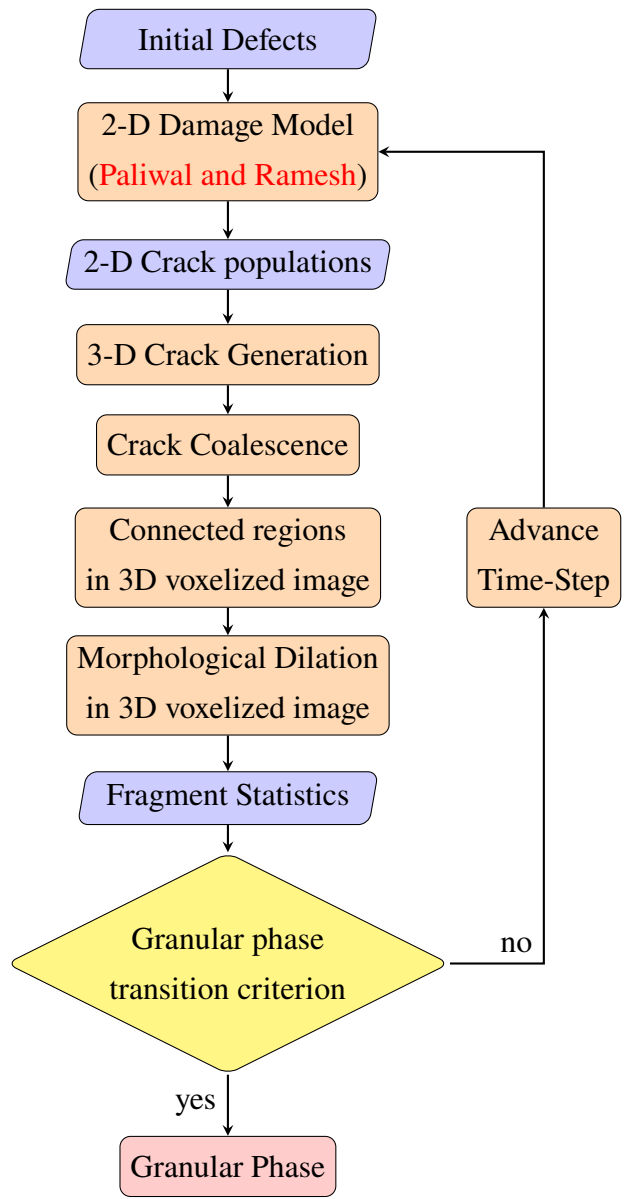
\begin{figure}[h!]
\centering
\begin{tikzpicture}[node distance=1.2cm, every text node part/.style={align=center}]
\node (in1) [io] {Initial Defects};
\node (pro1) [process, below of=in1, minimum height=1.2cm, yshift=-0.3cm] {2-D Damage Model\\(\citeauthor{PALIWAL2008})};
\node (out1) [io, below of=pro1, yshift=-0.3cm] {2-D Crack populations};
\node (pro2) [process, below of=out1] {3-D Crack Generation};
\node (pro3) [process, below of=pro2] {Crack Coalescence};
\node (pro4) [process, below of=pro3, yshift=-0.3cm] {Connected regions\\ in 3D voxelized image};
\node (pro5) [process, below of=pro4, yshift=-0.5cm] {Morphological Dilation\\ in 3D voxelized image};
\node (out2) [io, below of=pro5, yshift=-0.3cm] {Fragment Statistics};
\node (dec1) [decision, below of=out2, yshift=-1cm] {Granular phase \\ transition criterion};
\node (pro6) [process, right of=pro4, xshift=3cm] {Advance \\ Time-Step};
\node (stop) [startstop, below of=dec1, yshift=-1.5cm] {Granular Phase};
\draw [arrow] (in1) -- (pro1);
\draw [arrow] (pro1) -- (out1);
\draw [arrow] (out1) -- (pro2);
\draw [arrow] (pro2) -- (pro3);
\draw [arrow] (pro3) -- (pro4);
\draw [arrow] (pro4) -- (pro5);
\draw [arrow] (pro5) -- (out2);
\draw [arrow] (out2) -- (dec1);
\draw [arrow] (dec1) --  node[anchor=east] {yes} (stop);
\draw [arrow] (dec1) -|  node[anchor=south east] {no} (pro6);
\draw [arrow] (pro6) |- (pro1);
\end{tikzpicture}
\caption{Outline of steps Involved in the fragmentation model}
 \label{fig:FlowChart}
\end{figure}

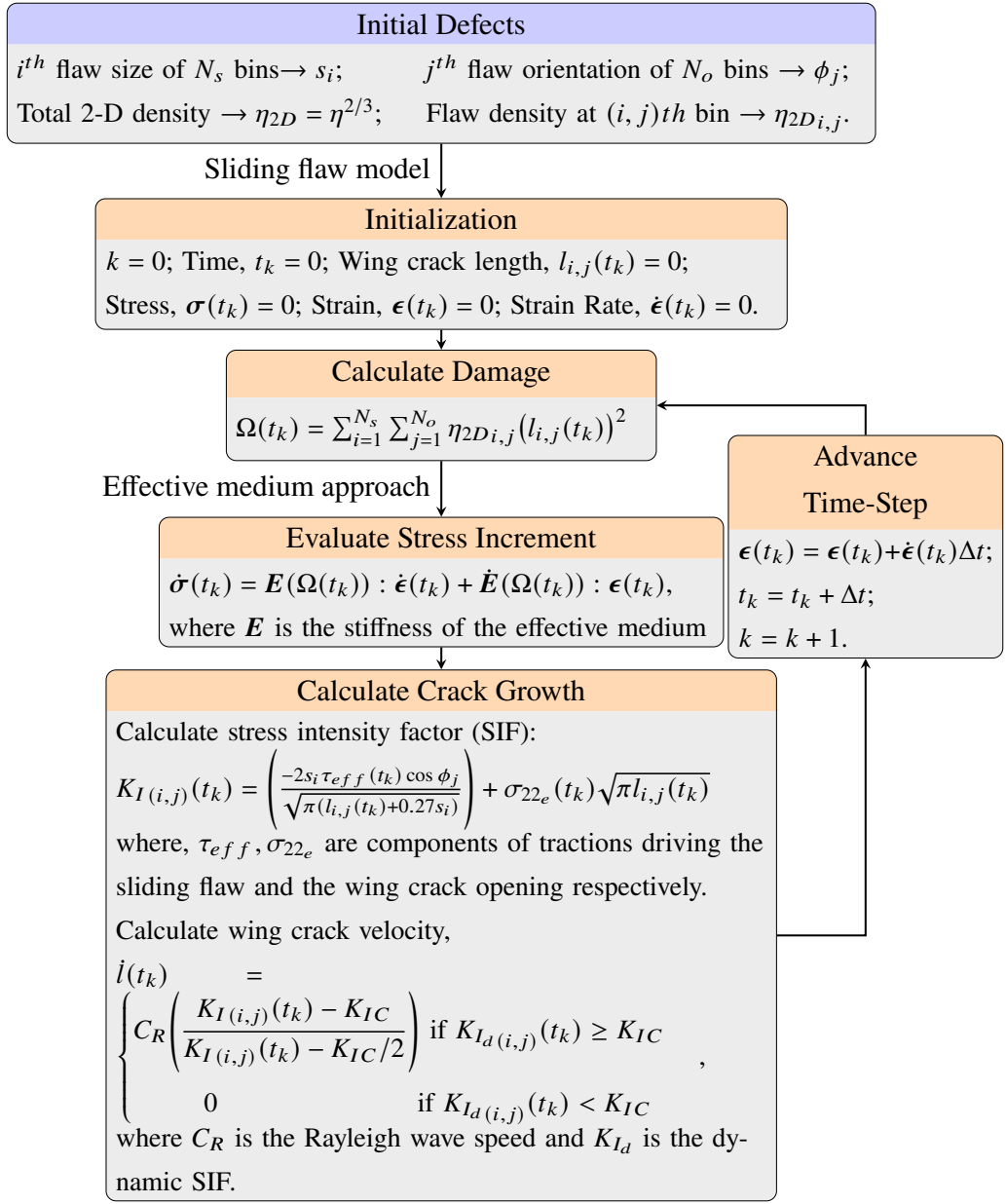
\begin{figure}[h!]
\centering
\resizebox{\textwidth}{!}{
\begin{tikzpicture}[node distance=3cm, every text node part/.style={align=center}]
\node (in1) [io1] {Initial Defects \nodepart[text width=12cm, font=\small]{two} $i^{th}$ flaw size of $N_s$ bins$\rightarrow s_i$;\hspace{1cm} $j^{th}$ flaw orientation of $N_o$ bins $\rightarrow \phi_j $;\\ Total 2-D density $\rightarrow \eta_{2D}=\eta^{2/3}$;\hspace{0.5cm} Flaw density at $(i,j)^{}th$ bin $\rightarrow {\eta_{2D}}_{i,j}$.};
\node (pro1) [process1, below of=in1,yshift=0.3cm] {Initialization  \nodepart[text width=9.5cm, font=\small]{two} $k=0$; Time, $t_k=0$; Wing crack length, $l_{i,j}(t_k)=0$;\\ Stress, $\boldsymbol{\sigma}(t_k)=0$; Strain, $\boldsymbol{\epsilon}(t_k)=0$; Strain Rate, $\dot{\boldsymbol{\epsilon}}(t_k)=0$.};
\node (pro2) [process1, below of=pro1,yshift=1cm] {Calculate Damage\nodepart[text width=5.8cm, font=\small]{two} $\Omega (t_k)=\sum_{i=1}^{N_s} \sum_{j=1}^{N_o} {\eta_{2D}}_{i,j} {\big(l_{i,j}(t_k)\big)}^2$};
\node (pro3) [process1, below of=pro2,yshift=0.5cm] {Evaluate Stress Increment\nodepart[text width=7.7cm, font=\small]{two} $\dot{\boldsymbol{\sigma}}(t_k)=\boldsymbol{E}(\Omega (t_k)):\dot{\boldsymbol{\epsilon}}(t_k)+\dot{\boldsymbol{E}}(\Omega (t_k)):\boldsymbol{\epsilon}(t_k)$,\\ where $\boldsymbol{E}$ is the stiffness of the effective medium};
\node (pro4) [process1, below of=pro3,yshift=-2cm] {Calculate Crack Growth\nodepart[text width=9.2cm, font=\small]{two} Calculate stress intensity factor (SIF):\\${K_I}_{(i,j)}(t_k)=\Bigg(\frac{-2s_i {\tau}_{eff}(t_k)\cos{\phi_j}}{\sqrt{\pi (l_{i,j}(t_k)+0.27s_i)}}\Bigg)+\sigma_{22_e}(t_k)\sqrt{\pi l_{i,j}(t_k)}$\\ where, ${\tau}_{eff},\sigma_{22_e}$ are components of tractions driving the sliding flaw and the wing crack opening respectively.\\ 
Calculate wing crack velocity,\\
${\dot{l}}(t_k)=\begin{cases} \!\begin{aligned} C_R\Bigg(\frac{{K_I}_{(i,j)}(t_k)-K_{IC}}{{K_I}_{(i,j)}(t_k)-K_{IC}/2}\Bigg) \mbox{ if   } {K_{I_d}}_{(i,j)}(t_k)\geq K_{IC} \end{aligned} \\
\!\begin{aligned}\hspace{1cm}0 \hspace{2.7cm}\mbox{ if   } {K_{I_d}}_{(i,j)}(t_k)< K_{IC}\end{aligned} \end{cases},$ \\ where $C_R$ is the Rayleigh wave speed and $K_{I_d}$ is the dynamic SIF.};
\node (pro5) [process1, right of=pro3, xshift=3cm, yshift=0.5cm] {Advance \\ Time-Step \nodepart[text width=3.6cm, font=\small]{two} $\boldsymbol{\epsilon}(t_k)=\boldsymbol{\epsilon}(t_k)+\dot{\boldsymbol{\epsilon}}(t_k)\Delta t$;\\$t_k=t_k+\Delta t$;\\$k=k+1$.};
\draw [arrow] (in1) -- node[anchor=east] {Sliding flaw model}(pro1);
\draw [arrow] (pro1) -- (pro2);
\draw [arrow] (pro2) -- node[anchor=east] {Effective medium approach}(pro3);
\draw [arrow] (pro3) -- (pro4);
\draw [arrow] (pro4) -| (pro5);
\draw [arrow] (pro5) |- (pro2);
\end{tikzpicture}}
\caption{Outline of wing crack growth based damage model (Paliwal $\&$ Ramesh 2008)}
 \label{fig:PR_Model}
\end{figure}
\FloatBarrier
\subsection{Initial defects and the crack growth model} \label{ssec:CrackSizeStat}
The analysis begins with randomly spaced initial defects, with a given size and orientation distribution. Two different defect size distributions are considered: (a) a delta distribution (i.e. a single fixed defect size) and (b) a lognormal distribution. The initial orientation distribution is taken to be a uniform distribution in the range $[0,2\pi]$ radians. Crack growth is modelled by a modified version of the \citeauthor{PALIWAL2008} model that accounts for orientation distribution of defects. The initial defect population is binned into a set of representative defect sizes and defect orientations.  \shortciteA{PALIWAL2008} calculate the growth of wing cracks associated with each crack population bin at a particular time instant (Fig: \ref{fig:PR_Model}). This model uses a discretized measure of a 2-dimensional scalar damage value, $\Omega$, to estimate the degradation of elastic properties and the resulting stress state using an effective medium approach. The scalar damage is defined as:
\begin{equation}
\label{eq:Damage2D}
\Omega=\int_{0}^{L_w} \eta_{2D}(l_w) l_w^2 f_{l_w}(l_w) dl_w,
\end{equation}
where $l_w$ is the half wing-crack length,  $\eta_{2D}(l_w)$ is the 2D crack density (number of cracks per unit area) at a given $l_w$, $L_w$ is the length of the largest wing crack.\par
\begin{figure}[h!]
\centering
 \includegraphics[width=0.5\textwidth]{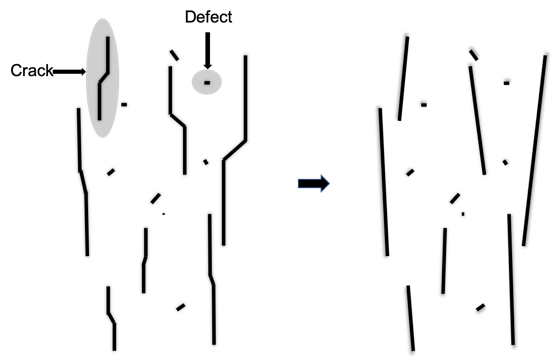}
 \caption{Simplification of defect wing crack assemblies to slanted cracks}
 \label{fig:slantedCracks}
 \end{figure}

Incremental crack growth is estimated from a dynamic crack growth criterion. Crack growth occurs when the dynamic stress intensity factor equates or exceeds the fracture toughness. In the fragmentation model, the crack lengths and orientations at a given stage are estimated from the wing crack lengths obtained using the \citeauthor{PALIWAL2008} model, by joining the tips of the wing cracks (Fig: \ref{fig:slantedCracks}). Unless otherwise specified, the material properties and model parameters in Table: \ref{tab:table1} have been used. 

\subsection{Simulation of three-dimensional cracks}\label{ssec:CrackSim3D}
Because the \citeauthor{PALIWAL2008} model is framed in two dimensions, the line cracks predicted by the model are translated to elliptical cracks in the three dimensional fragmentation model. The cracks are simulated in a three dimensional box with periodic geometry. This simulation box is a collection of cuboid voxels, and the size of each voxel is referred to as the resolution size. A voxel either belongs to a crack, or it is part of the intact material. The major axis of an elliptical crack is the crack length and its inclination with respect to the y-axis is the effective crack orientation from the \citeauthor{PALIWAL2008} model. The aspect ratio of the cracks has been chosen to be 1:1, but it can be set to any value. The size of the simulated sample should be large enough to accurately capture a representative range of crack sizes, including the largest cracks. The resolution size should be small enough to capture the smallest crack sizes and the corresponding small fragments. Of course, larger simulation boxes and finer resolution lead to increased computational effort. \par
\begin{figure}
\centering
 \includegraphics[width=0.75\textwidth]{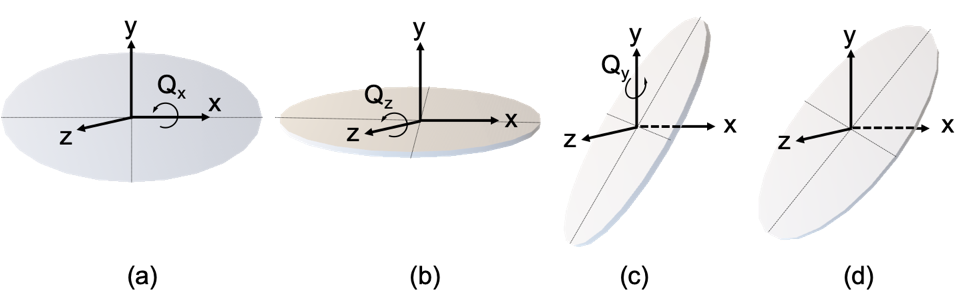}
 \caption{3-D elliptical crack generation}
 \label{fig:3DCracks}
 \end{figure}

For the 3D problem, let the y-axis refer to the direction of maximum principal compression. The angle of inclination of the major axis of an ellipse with the y-axis is the same as the complement of the corresponding 2D crack orientation ($\theta_{2D}$). In addition, the projection of the major axis on the plane perpendicular to direction of maximum principal compression (or the xz plane) is random. Now given a certain fixed major axis orientation, the minor axis can lie on any plane containing the major axis. This ensures that the only constraint we apply on the ellipses is the inclination of its major axis with respect to the direction of maximum principal compression (y-axis). This has been accomplished by generating an ellipse in the xy plane with its major axis aligned along the x-axis, and then rotating it first by a random angle about the x-axis ($\theta_x$), followed by the corresponding 2D crack orientation angle about the z-axis ($\theta_z=\theta_{2D}$), and then a random angle about the y-axis ($\theta_y$). The steps have been shown in Fig: \ref{fig:3DCracks}. The corresponding rotation matrices are $Q_x$, $Q_z$, $Q_y$ in the order of rotation. It can be shown that these set of rotations are equivalent to generating an ellipse with a major axis inclined at $\pi/2-\theta_{2D}$ with the y-axis and its projection having a random angle in the xz plane, followed by rotating that ellipse about its major axis by a random angle.

\subsection{Crack coalescence due to crack growth} \label{ssec:CrackCoa}

As cracks grow under increased loading, they are more likely to coalesce with neighboring cracks. \shortciteA{HUQ2019} has developed a probabilistic two dimensional crack coalescence model for fixed flaw orientation. Coalescence of three-dimensional cracks, however, is a complicated phenomenon. Different modes of coalescence have been recorded in the literature \shortcite{WONG1998}, and the competition between cracks growing in a competitive environment means that they would prefer one crack over the other to coalesce with \shortcite{Wong2001}. Multiple factors like the length and orientation of individual cracks and the crack bridges affect the order and mode of crack coalescence. The \citeauthor{PALIWAL2008} model calculates the instantaneous growth of individual crack populations growing in isolation in an effective matrix, but it does not address coalescence. Instead, this coalescence is addressed in the 3D fragmentation model presented here, which considers size, proximity and orientation of cracks to determine if coalescence occurs. While there is some crack coalescence that occurs throughout loading, crack coalescence accelerates dramatically when the cracks are significantly large and comminution is about to begin. Since the objective of this work is focused on the onset of comminution, when rapid coalescence leads to fragmentation, crack coalescence is assumed to be instantaneous and the number density of individual crack populations due to coalescence is assumed to remain constant. The following section describes an analytical model for 2-D crack coalescence followed by two numerical approaches of implementing it in three dimensions and identifying the onset of fragmentation.
 
 \subsubsection{Calculation of stress intensity factor for crack coalescence}\label{sssec:CrackGrowth}

\begin{figure}
\centering
 \includegraphics[width=0.75\textwidth]{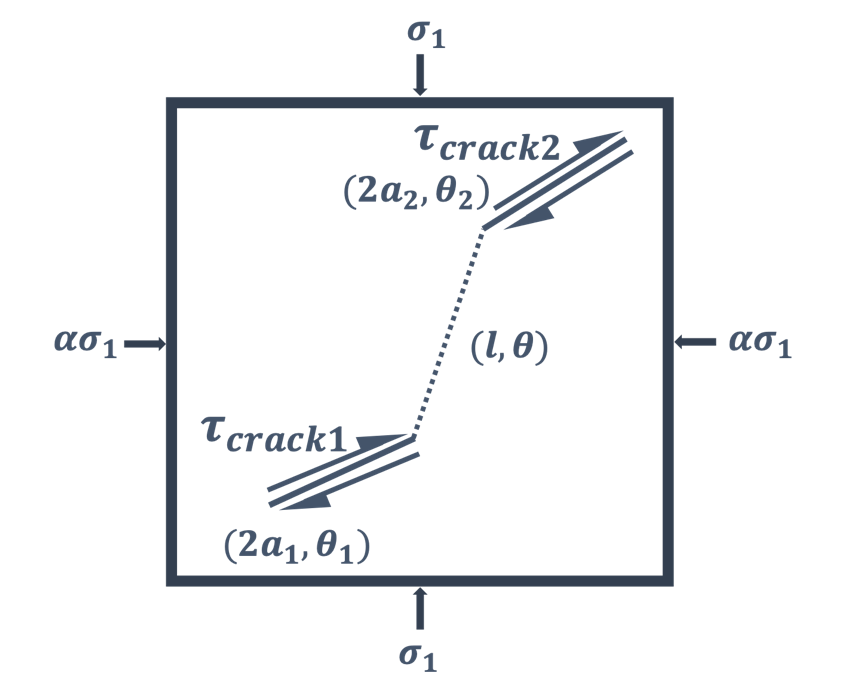}
 \caption{2-D crack problem geometry }
 \label{fig:SIF_2D}
 \end{figure}

Calculating the stress intensity factor for crack growth in a three-dimensional problem involves a complete 3D stress analysis. Given the multiple possible elliptical crack orientations in the current problem, this might be quite challenging as well as expensive. For simplification, the problem is approached by a two dimensional model (Fig: \ref{fig:SIF_2D}). Even in this simplified representation, there are two cracks which are associated with orientation and length. The crack bridge that joins the two cracks also has an orientation and length. This six-parameter problem has to be solved numerically for all possible position and orientation scenarios, which is still largely infeasible. In order to avoid such expensive numerical calculations, the problem is simplified by assuming that the stress field acting on a crack is a combination of the global stresses that act on the crack face as well as the stresses acting on it due to the stress field of the nearby crack. It is worth mentioning that compressive stresses acting on the crack face do not contribute significantly to crack growth and/or coalescence, so this analysis focuses on shear stresses. Therefore, the stresses acting on a crack are as follows:
\noindent
\begin{enumerate}
\item Shear stresses acting along crack 1:
\begin{enumerate}
    \item Global stresses, 
  \begin{equation} 
  \label{eq:G_ShearStress_C1}
\tau_{crack1}^{global}=\frac{\sigma_1}{2}(1-\alpha)\sin{(2\theta_1)};
\end{equation}
  \item Stresses from crack 2, $\tau_{crack1}^{c2}$.
\end{enumerate}
\item Shear stresses acting along crack 2:
\begin{enumerate}
  \item Global stresses,
  \begin{equation} 
  \label{eq:G_ShearStress_C2}
\tau_{crack2}^{global}=\frac{\sigma_1}{2}(1-\alpha)\sin{(2\theta_2)};
\end{equation}
  \item Stresses from crack 1, $\tau_{crack2}^{c1}$ .
\end{enumerate}
\end{enumerate}

\noindent
The stress on crack 1 is related to the stresses on crack 2, which in turn relates to the stresses on crack 1 (from 2(b) above), creating a feedback loop in the analysis. Assuming that the stresses on crack 2 due to crack 1 (2(b) above) are smaller than those due to the global stress (2(a) above), we can ignore the effects of crack 1 on crack 2, when calculating the effects of crack 2 on crack 1 (1(b) above).
\noindent
From \shortciteA{HELLO2012}, for a pure mode II isolated crack of length $2a$ with an infinite boundary defined in a complex plane with the origin at the crack center by complex number $z$ and polar coordinates at the crack tip $(r,\theta)$,
\begin{equation} 
  \label{eq:Hello_Mode2}
\sigma_{ij}^2 (r,\theta)=\begin{cases} \!\begin{aligned} \sigma_{12}^{\infty} \sum_{n=0}^{\infty}\Big[\psi_n^{'} (a) g_n^{2,ij} (\theta) r^{n-1/2}\Big] \hspace{1cm} \mbox{ for |z-a|<2a}\end{aligned} \\
\!\begin{aligned}\sigma_{12}^{\infty} \sum_{n=0}^{\infty}\Big[\tilde{\psi}_n^{'} (a) \tilde{g}_n^{2,ij} (\theta) r^{-n}\Big]  \hspace{1.5cm} \mbox{ for |z-a|>2a}\end{aligned}\end{cases} ,
\end{equation}
where,
\begin{subequations} 
  \label{eq:Hello_Param}
  \begin{equation}
  \label{eq:Hello_Param_1}
\psi_n^{'} (a)=\frac{(-1)^{n+1} (2n+1)!}{2^{3n+1/2} (2n-1) (n!)^2 a^{n-1/2} },
  \end{equation}
  \begin{equation}
  \label{eq:Hello_Param_2}
g_n^{2,11} (\theta)=1/2\Big[(n+7/2)  \sin{((n-1/2)\theta)}-(n-1/2)  \sin{((n-5/2)\theta)} \Big],
  \end{equation}
  \begin{equation}
  \label{eq:Hello_Param_3}
g_n^{2,22} (\theta)=1/2\Big[(-n+1/2)  \sin{((n-1/2)\theta)}+(n-1/2) \sin{((n-5/2)\theta)} \Big],
  \end{equation}
  \begin{equation}
  \label{eq:Hello_Param_4}
g_n^{2,12} (\theta)=1/2\Big[(n+3/2)  \cos{((n-1/2)\theta)}-(n-1/2) \cos{((n-5/2)\theta)} \Big],
  \end{equation}
  \begin{equation}
  \label{eq:Hello_Param_5}
\tilde{\psi}_n^{'} (a)=\frac{(-1)^n (n-1)(2n)!a^n}{2^n (2n-1) (n!)^2 },
  \end{equation}
  \begin{equation}
  \label{eq:Hello_Param_6}
\tilde{g}_n^{2,11} (\theta)=\Big[(n/2-2)  \sin{(n\theta)}+(-n/2)  \sin{((n+2)\theta)} \Big],
  \end{equation}
  \begin{equation}
  \label{eq:Hello_Param_7}
\tilde{g}_n^{2,22} (\theta)=\Big[(-n/2)   \sin{(n\theta)}+(n/2)  \sin{((n+2)\theta)} \Big],
  \end{equation}
  \begin{equation}
  \label{eq:Hello_Param_8}
\tilde{g}_n^{2,12} (\theta)=\Big[(-n/2+1)  \cos{(n\theta)}+(n/2)  \cos{((n+2)\theta)} \Big].
  \end{equation}
\end{subequations}

The complete asymptotic stress field from \shortciteA{HELLO2012} has been used to determine the shear stress contribution from crack 2 on crack 1.

\begin{figure}
\centering
 \includegraphics[width=0.5\textwidth]{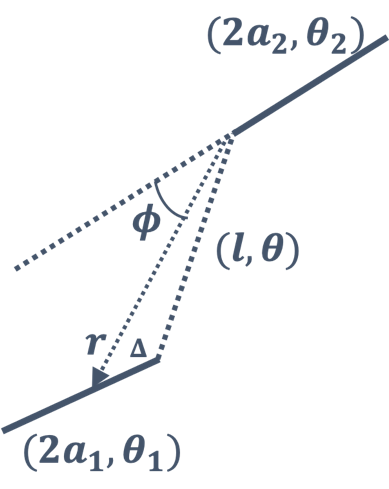}
 \caption{2-D crack problem geometry }
 \label{fig:Geo_2D}
 \end{figure} 
 
From the geometry of the problem (Fig: \ref{fig:Geo_2D}),
\begin{subequations} 
  \label{eq:Crack_2D_Geo}
\begin{equation} 
  \label{eq:Crack_2D_Geo_phi}
\phi=\arctan{\Bigg[\frac{l\sin{(\theta-\theta_2)}-\Delta \sin{(\theta_2-\theta_1)} }{l \cos{(\theta-\theta_2)}+\Delta\cos{(\theta_2-\theta_1)}}\Bigg]},
\end{equation}
\begin{equation} 
  \label{eq:Crack_2D_Geo_r}
r=\Delta\cos{(\theta_2-\theta_1+\phi)}+l\cos{(\theta-\theta_2-\phi)}.
\end{equation}
\end{subequations}

Using Eq: \ref{eq:Hello_Mode2} and transforming the stresses to obtain the shear stresses along crack 1 orientation, the shear component along crack 1 due to crack 2 stresses is given as: 

\begin{equation} 
  \label{eq:Tau_C1_dueC2}
\tau_{crack1}^{c2}=\frac{\tau_{crack2}^{global}}{2a_1}\int_0^{2a_1}\kappa(a_2,r,\phi,\zeta)d\Delta,
\end{equation}
where,
\begin{equation} 
  \label{eq:Kappa_Tau_C1_dueC2}
\kappa(a_2,r,\phi,\zeta)=\begin{cases} \!\begin{aligned} &\sum_{n=0}^{\infty} \psi_n^{'}(a_2)r^{n-1/2}\big[(n+3/2)\cos{((n-1/2)\phi+2\zeta)}\\ & -(n-1/2)\cos{((n-5/2)\phi+2\zeta)}\big] & \mbox{when  } r\leq 2a_2 \end{aligned} \\ 
\!\begin{aligned} & \sum_{n=0}^{\infty} \tilde{\psi}_n^{'}(a_2)r^{-n}\big[(1-n/2)\cos{(n\phi-2\zeta)}\\ & +n/2\cos{((n+2)\phi-2\zeta)}\big] & \hspace{1.8cm} \mbox{when  } r> 2a_2 \end{aligned} \end{cases} ,
\end{equation}
\noindent
and, $\zeta=-(\theta_2-\theta_1+\phi)$.\\
The total shear stress acting on crack 1 is $\tau_{crack1}=\tau_{crack1}^{global}+\tau_{crack1}^{c2}$.\\
Therefore,
\begin{equation} 
  \label{eq:Tau_C1}
\tau_{crack1}=\frac{\sigma_1}{2} (1-\alpha)\big[\sin{(2\theta_1)}+\sin{(2\theta_2)} \frac{1}{2a_1}\int_0^{2a_1}\kappa(a_2,r,\phi,\zeta)d\Delta \big].
\end{equation}
Crack growth will occur when the mode I stress intensity factor (SIF), $K_I$ along the crack bridge exceeds the critical stress intensity factor ($K_{IC}$) for the material.
\begin{equation} 
  \label{eq:CrackGrowthCrit}
K_I \geq  K_{IC} \\
\implies \tau_{crack1}\sqrt{\pi a_1}f_I(\theta-\theta_1) \geq  K_{IC},
\end{equation}
where, $f_I(\lambda)=-\frac{3}{4}\sin{(\frac{3\lambda}{2})}-\frac{3}{4}\sin{(\frac{\lambda}{2})}$.\\
We can also similarly use a mixed mode crack growth criterion based on energy release rate. For the current work however, Eq: \ref{eq:CrackGrowthCrit} has been used for crack coalescence.\par
Extending the crack coalescence approach to three dimensions involves dealing with complicated geometries and locations of multiple elliptical cracks as well as accounting for three dimensional stresses. There are no known analytical solutions for such problems. In this paper two approaches to tackle three dimensional crack coalescence are highlighted. In one of the approaches, pairs of individual cracks are connected by coalescence surfaces running along the crack edges (Sec: \ref{sssec:CoaSurfApp}), while in the other approach each crack edge is surrounded by a probable zone along which coalescence is likely to occur (Sec: \ref{sssec:CoaZoneApp}). In either case, the three dimensional problem has been simplified to a two dimensional problem similar to the one discussed above, where the 2-D orientation of cracks and crack bridge are equal to the inclination of the corresponding 3-D feature with the y-axis (direction of maximum principal compression). The lengths in the 2-D problem are simply the lengths of the corresponding features in the 3-D problem.

\subsubsection{Coalescence surface approach}\label{sssec:CoaSurfApp}

\begin{figure}
\centering
 \includegraphics[width=0.4\textwidth]{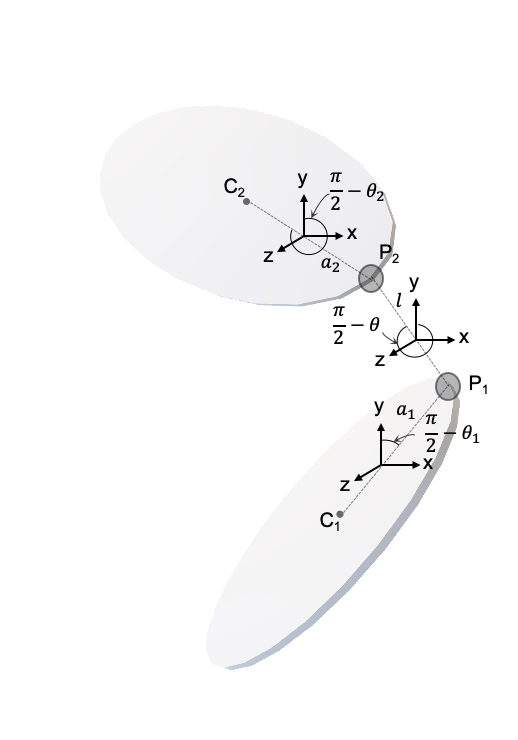}
 \caption{Estimating 2-D geometry from 3-D problem}
 \label{fig:Geo_2D_3D}
 \end{figure}

Fig: \ref{fig:Geo_2D_3D} illustrates the first approach for 3D crack coalescence. Consider a crack ($C_1$) and one of the voxels ($P_1$) along the edge of that particular crack, as well as the nearest voxel ($P_2$) lying along the edge of another crack ($C_2$). The approach in Sec: \ref{sssec:CrackGrowth} is used to determine whether crack growth is feasible along the direction $P_1P_2$ that does not exceed a threshold distance. If crack growth is not feasible, then the nearest voxel on another crack within the region is assessed. The feasibility of crack growth is again determined by using the 2-dimensional crack coalescence model in Sec: \ref{sssec:CrackGrowth}. The length and the complement of the inclination, relative to the direction of maximum principal compression (y-axis), of the line joining the center of the crack, $C_1$, and the crack edge point, $P_1$, are calculated as $a_1$ and $\theta_1$ respectively. Similarly, the corresponding values for the line joining the center of $C_2$ and $P_2$ are calculated as $a_2$  and $\theta_2$. The length and complement of the inclination, relative to the y-axis, of the crack bridge joining the points $P_1$ and $P_2$ are calculated as $l$  and $\theta$ (See Fig: \ref{fig:Geo_2D_3D}). If the mode I stress intensity factor ($K_I$) for the growth of crack $C_1$ along the bridge direction $P_1P_2$, exceeds the critical stress intensity factor (Eq: \ref{eq:CrackGrowthCrit}), then the crack will grow in that direction and coalescence would occur. \par
\begin{figure}[htpb]
  \centering
    \subfloat[]{\label{fig:CoaSA1}\includegraphics[width=0.45\textwidth]{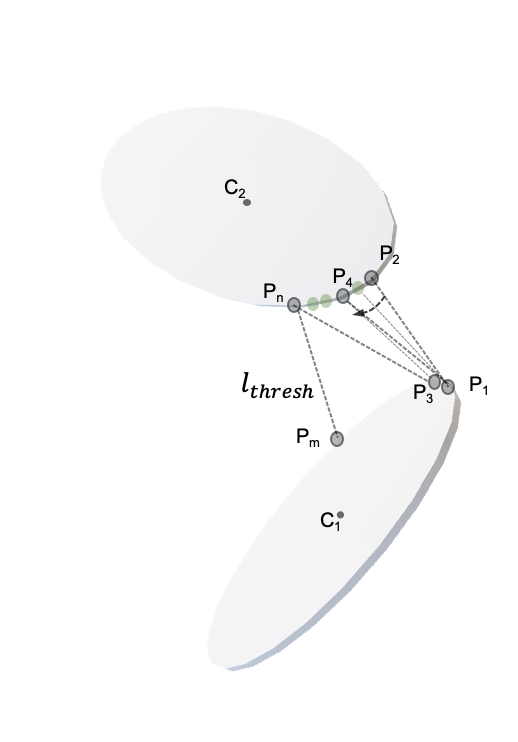}}\hfill
    \subfloat[]{\label{fig:CoaSA2}\includegraphics[width=0.55\textwidth]{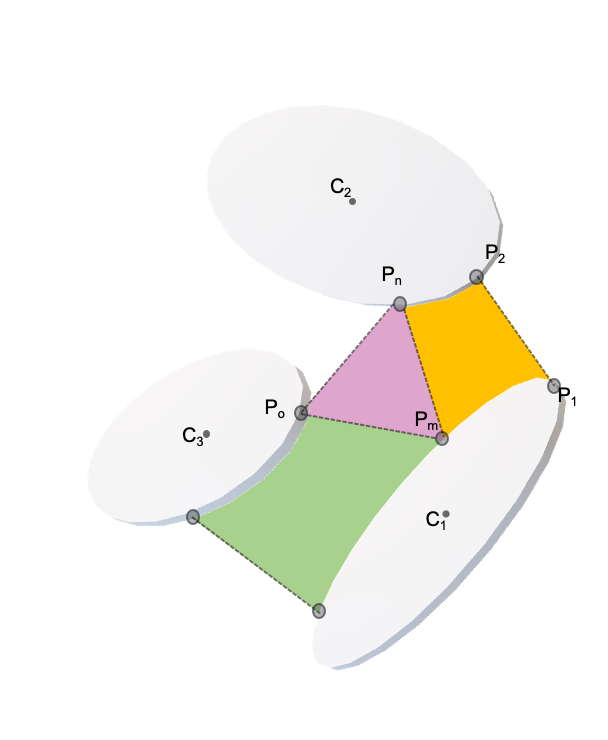}}\hfill 
  \caption{Steps involved in coalescence surface approach ($l_{thresh}$ denotes the threshold distance)}
  \label{fig:CoaSASteps}
\end{figure}

After $K_I$ is assessed for $P_1P_2$, the next step is to move to an adjacent voxel on the edge of crack $C_1$ and then identify the closest point along the edge of crack $C_2$, and repeat the analysis to determine if there is crack bridging. The direction on $C_2$ along which the distance to the adjoining point, $P_3$, on crack $C_1$ reduces is chosen. All voxels along the triangle $P_1P_2P_3$ are assigned to be cracked. After fixing the direction, we keep pairing points along $C_2$ with $P_1$ and assigning voxels containing incremental triangles to cracked regions, till we reach a point $P_4$ beyond which all points exceed a certain threshold distance. This threshold distance explained in Sec: \ref{sssec:ThreshDist} is the measure of potential crack growth over a given future time period. Now from $P_3$ again a similar search is done for points to pair on $C_2$ which are less than the threshold distance, along the aforementioned direction. This process of mapping points has been illustrated in Fig: \ref{fig:CoaSA1}. This process is repeated until we reach a point $P_m$ on $C_1$ and $P_n$ on $C_2$ beyond which no points can be found on $C_2$ that are less than the threshold distance. This is when we search for other cracks to pair with and repeat the same procedure with crack $C_3$ starting with $P_o$ the point closest to $P_m$ on $C_3$. After pairing these points, additional coalescence surfaces are created as shown on Fig: \ref{fig:CoaSA2} using the approach of incremental triangles. We also check whether the $P_o$ and $P_n$ are less than the threshold distance. If so, we end up creating a new surface enclosed in the triangle $P_mP_nP_o$ by assigning all voxels along the surface to cracked regions.

\subsubsection{Coalescence zone approach} \label{sssec:CoaZoneApp}

\begin{figure}[htpb]
  \centering
    \subfloat[Crack without coalescence zone]{\label{fig:Crack}\includegraphics[width=0.45\textwidth]{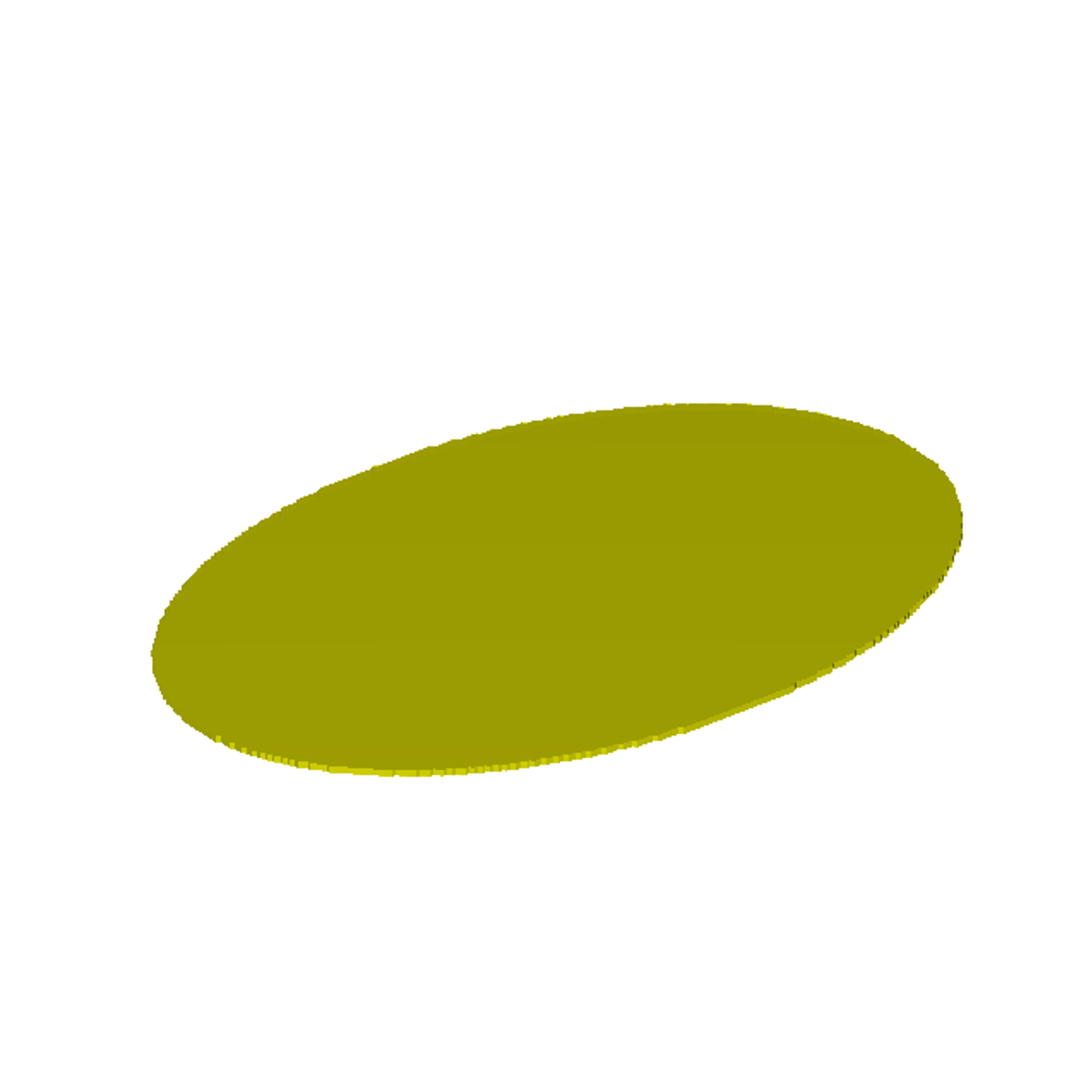}}\hfill
    \subfloat[Crack with coalescence zone]{\label{fig:CoaZoneK1}\includegraphics[width=0.45\textwidth]{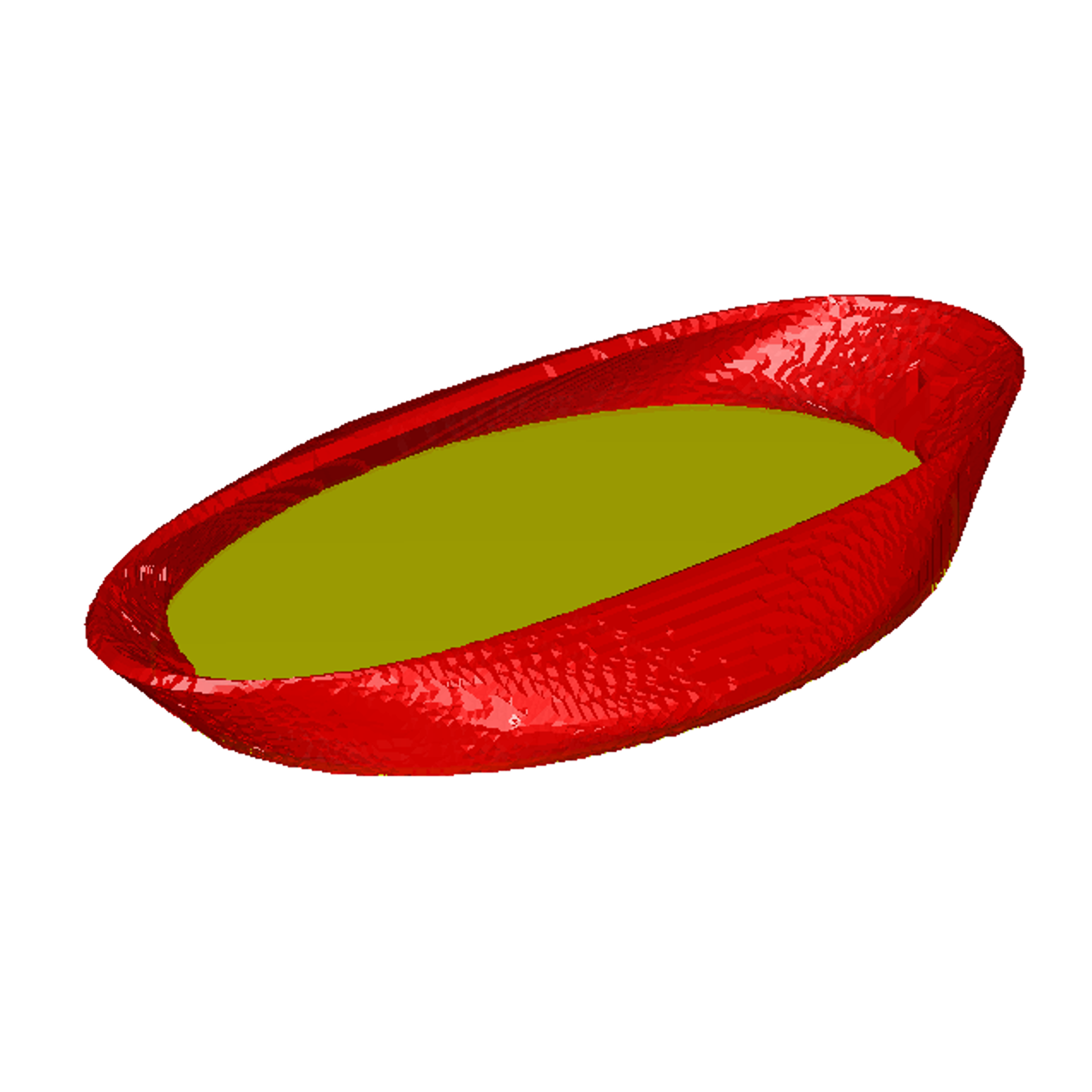}}\hfill 
  \caption{Representative image of an elliptical crack (in yellow) with a coalescence zone (in red)}
  \label{fig:Coa_Zone}
\end{figure}

A more efficient approach is to specify a coalescence zone surrounding the edge of a crack, across which coalescence can take place. The coalescence zone is a volume measure as opposed to a coalescence surface. In this case, for each crack edge point the average bridge length in a particular direction, for which crack coalescence is feasible, is determined by solving a 2-D crack coalescence problem similar to Sec: \ref{sssec:CrackGrowth}. For the 2-D problem in Fig: \ref{fig:SIF_2D}, the bridge length, $l$, is determined for a given value of $\theta_1$, $2a_1$, $\theta$ for all possible combinations of $\theta_2$ and $2a_2$. $2a_1$ is the major axis length of the crack $C_1$ and $\theta_1$ is the complement of the angle this segment makes with respect to the y-axis (Fig:  \ref{fig:Geo_2D_3D}). Similarly the values for $2a_2$, $\theta_2$, $l$ and $\theta$ are found. In the end, for each crack edge point we have a set of $l$ values corresponding to different  $\theta$ values for multiple combinations of $\theta_2$ and $2a_2$. After repeating the process for all points along the edge of a crack, we can specify a domain around the entire crack edge, where coalescence is feasible. The maximum length of crack coalescence is set as the threshold distance mentioned in section Sec: \ref{sssec:ThreshDist}. Fig: \ref{fig:Coa_Zone} shows the representative image of the coalescence zone (in red) around an elliptical crack (in yellow). Fig: \ref{fig:Crack} shows the crack without the coalescence zone and Fig: \ref{fig:CoaZoneK1} shows the crack with the coalescence zone.

\subsubsection{Threshold distance for crack coalescence}\label{sssec:ThreshDist}

\begin{figure}
\centering
 \includegraphics[width=0.75\textwidth]{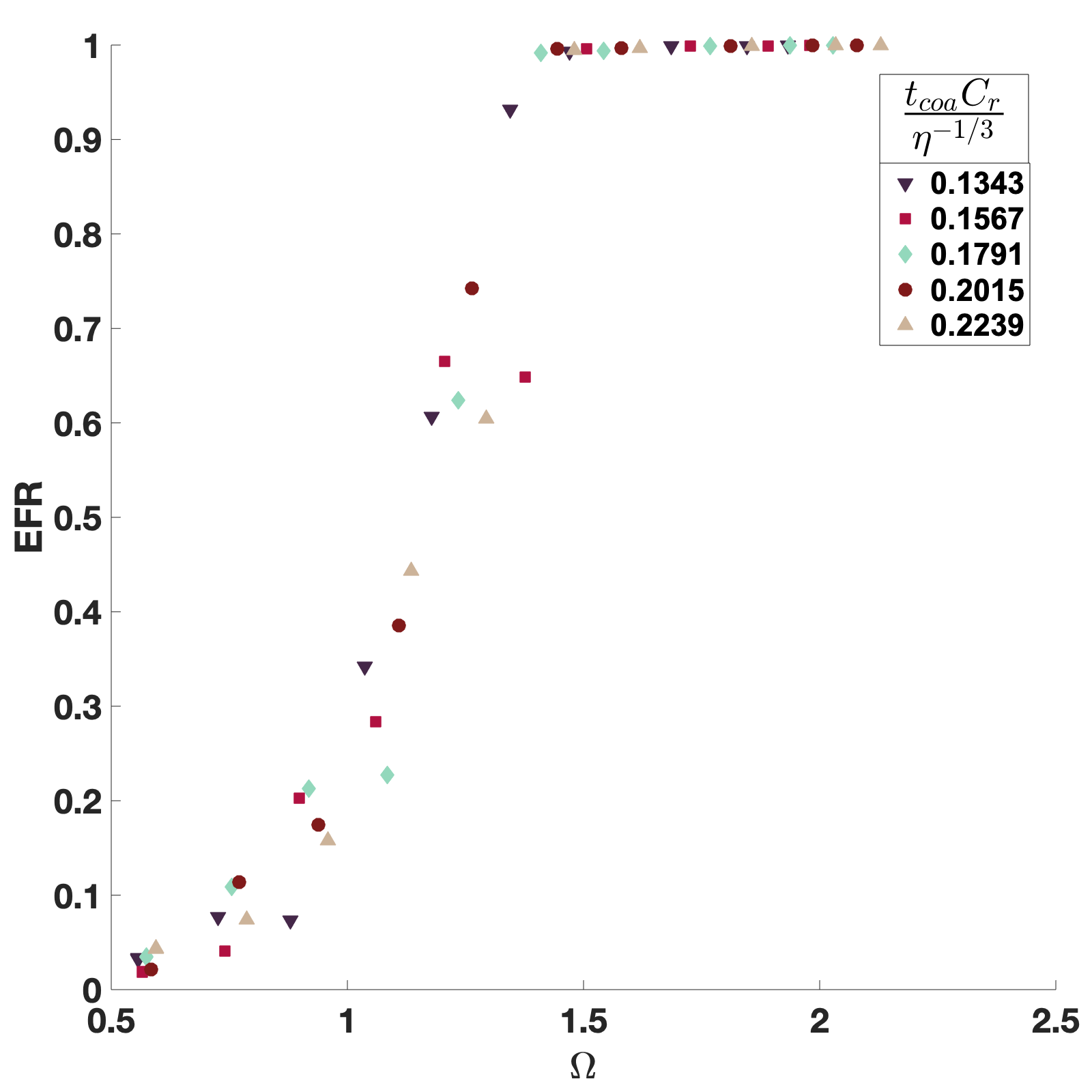}
 \caption{EFR evolution with damage with change in coalescence time}
 \label{fig:EFRdamageCoa}
 \end{figure}

The threshold distance is defined here as the maximum projected distance a crack can grow within a specific period of time. In the application to later stage crack coalescence, coalescence is allowed to occur across a given distance along favourable crack directions. In the wing crack growth problem, this distance is obtained by projecting the current crack velocity and acceleration into future timesteps and making sure that the crack velocity never exceeds the Rayleigh wave speed ($C_R$) of the material. For very high damage values, prior to fragmentation, it is difficult to determine the stress decay with increase in damage. At this stage the solid is heavily cracked and still not in its granular phase. This state of intensive cracking violates the dilute approximations used in the formulation of continuum-based damage models. To overcome this issue, it has been assumed that in the post peak strength phase of rapid damage growth, beyond a threshold damage value which determines the limit of applicability of continuum damage models, stress drops at a constant rate. The average stress, during the post-peak stress drop period, averaged over a given number of timesteps, is the stress that is used in the crack coalescence model, and the crack growth over these timesteps is the threshold or link distance in our model. If we use more timesteps, the stress drops but the allowable threshold distance increases, balancing each effect. This has been demonstrated in Fig: \ref{fig:EFRdamageCoa}. In the figure $t_{coa}$ is the time over which crack coalescence occurs, $C_R$ is the Rayleigh wave speed and $\eta^{-1/3}$ represents an average measure of crack center spacing. The change in effective fragmentation ratio (a measure of the degree of fragmentation in a given material and will be defined later in Sec: \ref{ssec:EFR}) with damage does not appear to be very sensitive to change in the number of time-steps used to calculate the threshold distance for crack coalescence. In our model we have chosen 5 timesteps for calculating the threshold distance. The damage values reported correspond to the total wing crack growth-based damage at the end of 5 timesteps.

\subsubsection{Comparison of the two approaches}\label{sssec:CoaCompApp}

The coalescence surface approach involves a search along the edge of each crack for nearby cracks, pairing points to connect with a new crack if they satisfy the growth criterion. The coalescence zone approach, on the other hand, creates a coalescence zone which is independent of the presence of nearby cracks and is based on an average measure of crack size and orientation. This makes the coalescence surface approach significantly more expensive than the coalescence zone approach. The coalescence surface approach connects only along crack edges, whereas the coalescence zone, although initiating new connections from a crack edge, can form connections anywhere on a crack. The coalescence surface approach connects with one crack edge at a time, while the zone approach can connect with multiple points on the edges of different cracks, starting from the same crack edge point. The coalescence zone approach is unable to resolve fragments smaller than the size of the coalescence zone. This is not a limitation of the coalescence surface approach, which can track longer crack bridge distances. The main limitation, however, lies in the computational expense of the coalescence surface approach. If multiple cracks are close to one another and along favorable directions of crack growth, one can expect all of them to eventually be interconnected, even though there might be a preferential order of crack connections. \par
\begin{figure}
 \centering
 \includegraphics[width=0.75\textwidth]{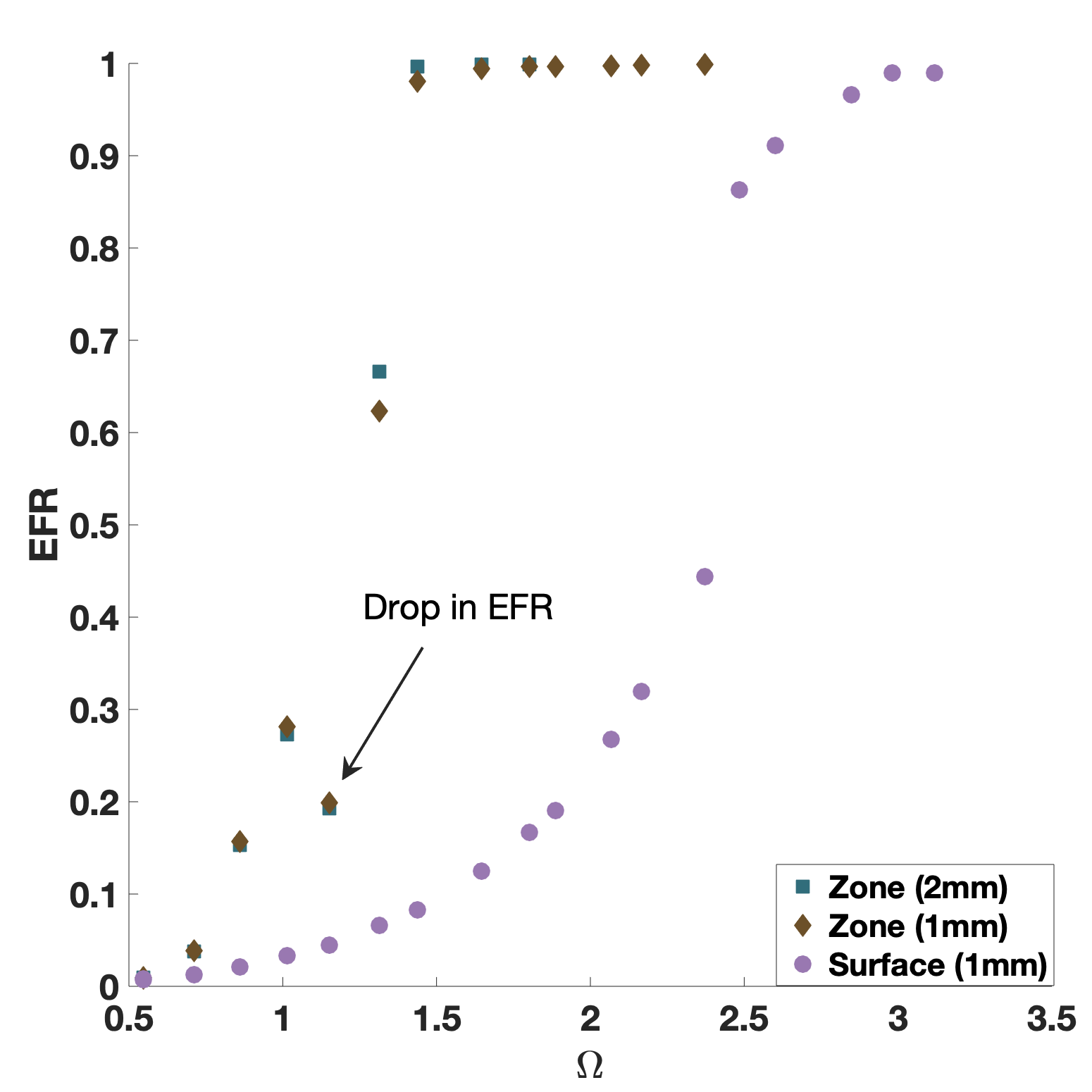}
 \caption{Comparison of the coalescence zone approach for 1mm and 2mm simulation box size and the coalescence surface approach for 1mm simulation box.}
 \label{fig:CompApp}
 \end{figure}

Fig: \ref{fig:CompApp} shows how the degree of fragmentation defined as the effective fragmentation ratio (further detail in \ref{ssec:EFR}) evolves with Damage for the two approaches and a smaller (1 mm) simulation box, along with a comparison of the coalescence zone approach for a larger simulation size (2 mm). As expected, the coalescence zone approach predicts a significantly higher degree of fragmentation than the coalescence surface approach. In view of this and given the significant computational expense of the coalescence surface approach, the coalescence zone approach has been used to model fragmentation for all simulations henceforth. We note that the coalescence zone approach often exhibits a sudden drop in EFR values (around a damage value of 1.2 in Fig: \ref{fig:S_E_D}) that is a consequence of the numerical approach will be discussed in Sec:  \ref{sec:Disc}. 
 
\subsection{Dilation and extracting fragment statistics}\label{sssec:DilFragStat}

Fig: \ref{fig:CompApp} shows convergence for different sample sizes using the coalescence zone approach. For all further analysis, we have used a 2 mm simulation box that ensures convergence. After fixing the size of the simulation box, the size of each voxel (also referred to as the resolution) is chosen depending on computational constraints and the ability to accurately model cracks. A coarser resolution means that we will not be able to resolve cracks and fragments smaller than the resolution size. The resolution size has been chosen such that it can model the smallest cracks in the system. For most of these simulations the average initial half flaw size is around 10 $\mu$. We have therefore chosen a cell size of 5 $\mu$ for all subsequent simulations.\par
\begin{figure}
 \includegraphics[width=\linewidth]{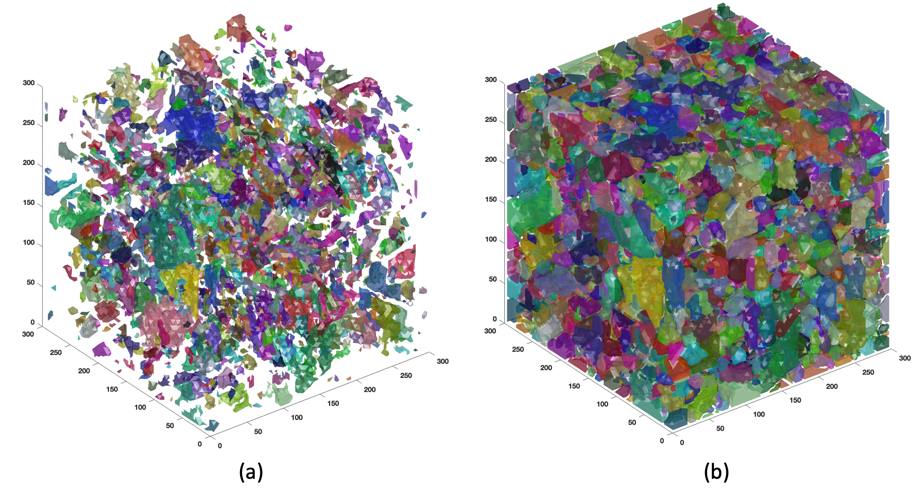}
 \caption{Representative image of connected regions (a) before and (b) after the dilation procedure}
 \label{fig:Con_Reg}
 \end{figure}

After simulating crack coalescence, the connected regions are obtained using MATLAB’s bwconncomp, assuming periodic boundaries. After finding the connected regions, a dilation procedure has been adopted to reallocate the voxels corresponding to coalescence zones and cracked regions to nearby connected regions (Fig: \ref{fig:Con_Reg}). This eliminates any loss in material volume that would arise as an artifact of our numerical approach. Finally, the fragment size statistics have been extracted from the dilated regions using the regionprops3 function. The fragment statistics that have been obtained include the fragment sizes, roundedness and solidity of fragments. The roundedness Index developed by \shortciteA{Hayakawa2005} is defined as $R=\frac{\mathcal{V}}{\mathcal{S}\sqrt[3]{abc}}$, where $\mathcal{V}$ and $\mathcal{S}$ are the volume and surface area of the object; parameters a,b,c are the principal axes of an equivalent ellipsoid. Solidity is defined as the volume fraction of voxels in the convex hull that are a part of the fragment. Roundedness Index value close to 0.33 would imply a perfectly round fragment, while solidity values lower than 1 signify increased angularity of fragments.

\section{Results - Transition to granular medium}

\begin{figure}[htpb]
  \centering
    \subfloat[Variation with $\Omega$]{\label{fig:SDS1}\includegraphics[width=0.5\textwidth]{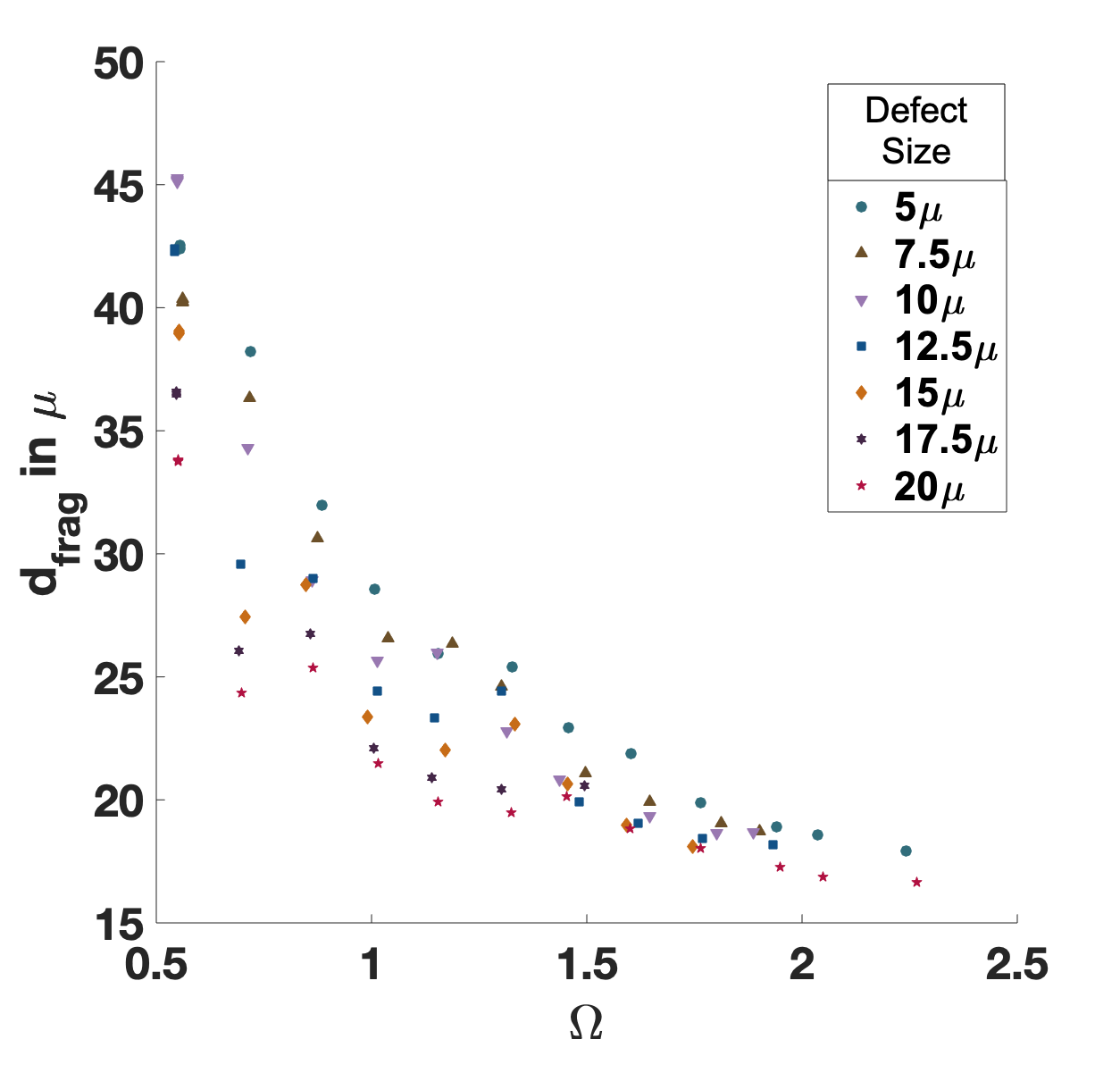}}\hfill
    \subfloat[Variation with EFR]{\label{fig:SDS2}\includegraphics[width=0.5\textwidth]{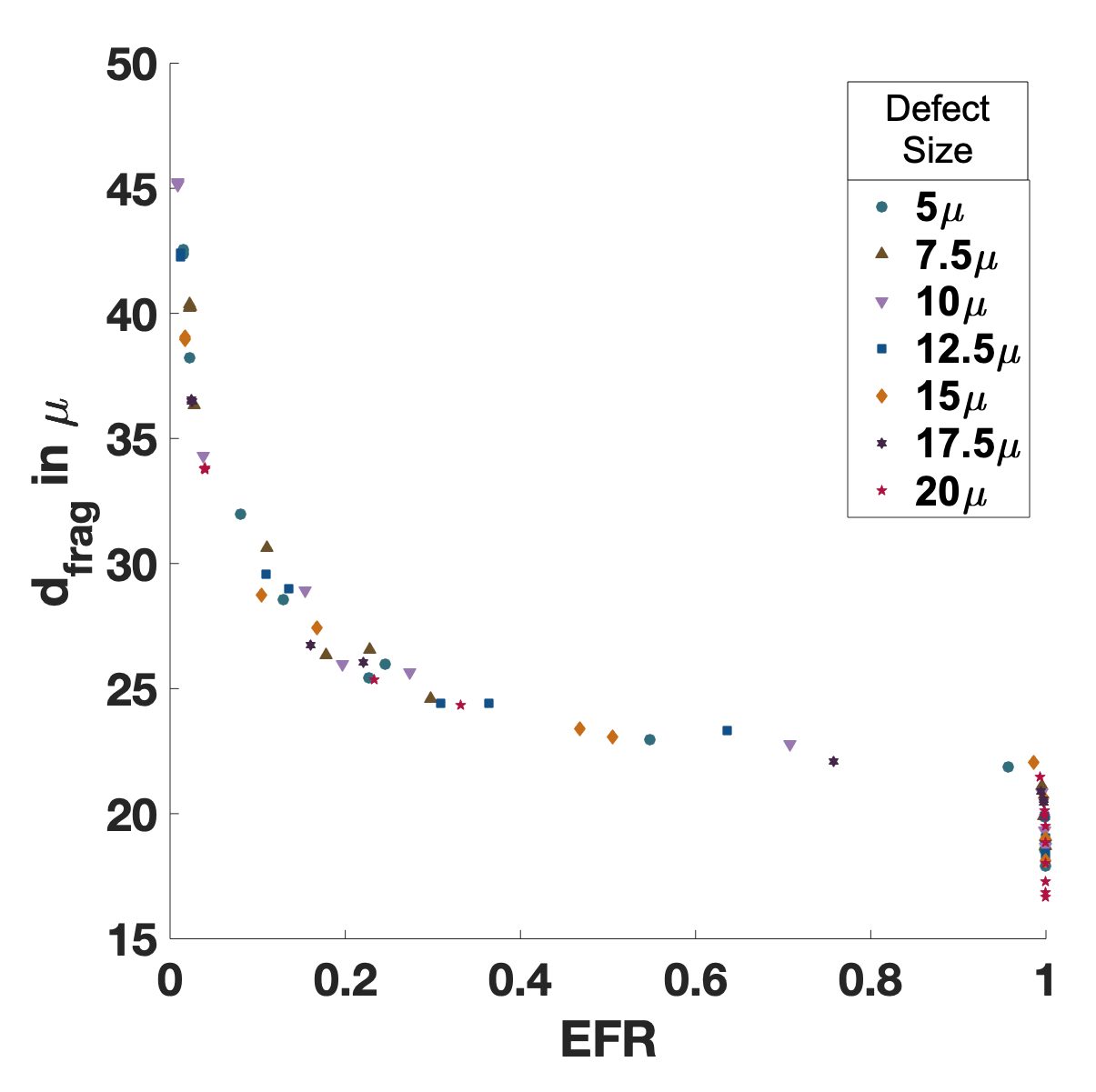}}\hfill 
  \caption{Variation of mean fragment size with damage and EFR at  $\eta=22x10^{12}\, cracks/m^3$ and $\dot{\epsilon}=10^6 s^{-1}$ for different initial defect sizes}
  \label{fig:SizeDefectSize}
\end{figure}

\subsection{Effective fragmentation ratio} \label{ssec:EFR}
\begin{figure}
\centering
 \includegraphics[width=0.75\textwidth]{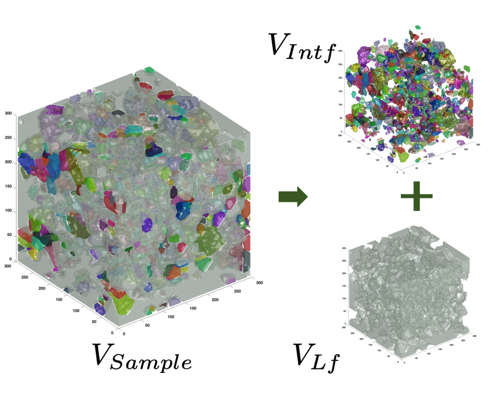}
 \caption{Effective fragmentation ratio, $EFR=1-V_{Lf}/V_{Sample}$}
 \label{fig:Efr}
 \end{figure}

\begin{figure}
\centering
 \includegraphics[width=\linewidth]{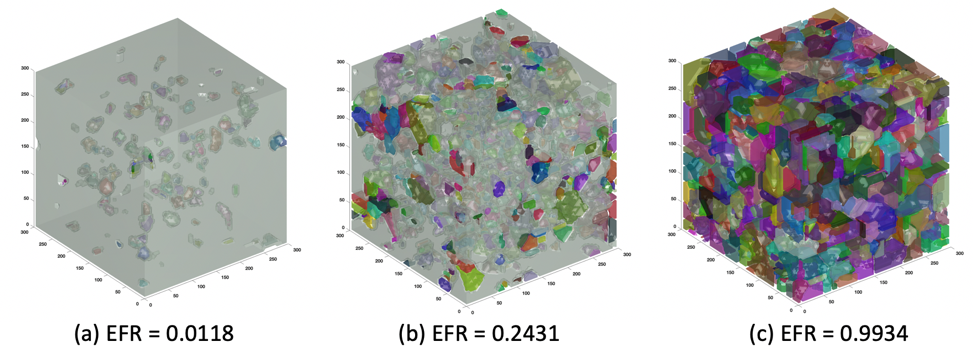}
 \caption{Fragmentation with increasing EFR}
 \label{fig:Efr_inc}
 \end{figure}

\begin{figure}
\centering
 \includegraphics[width=0.75\linewidth]{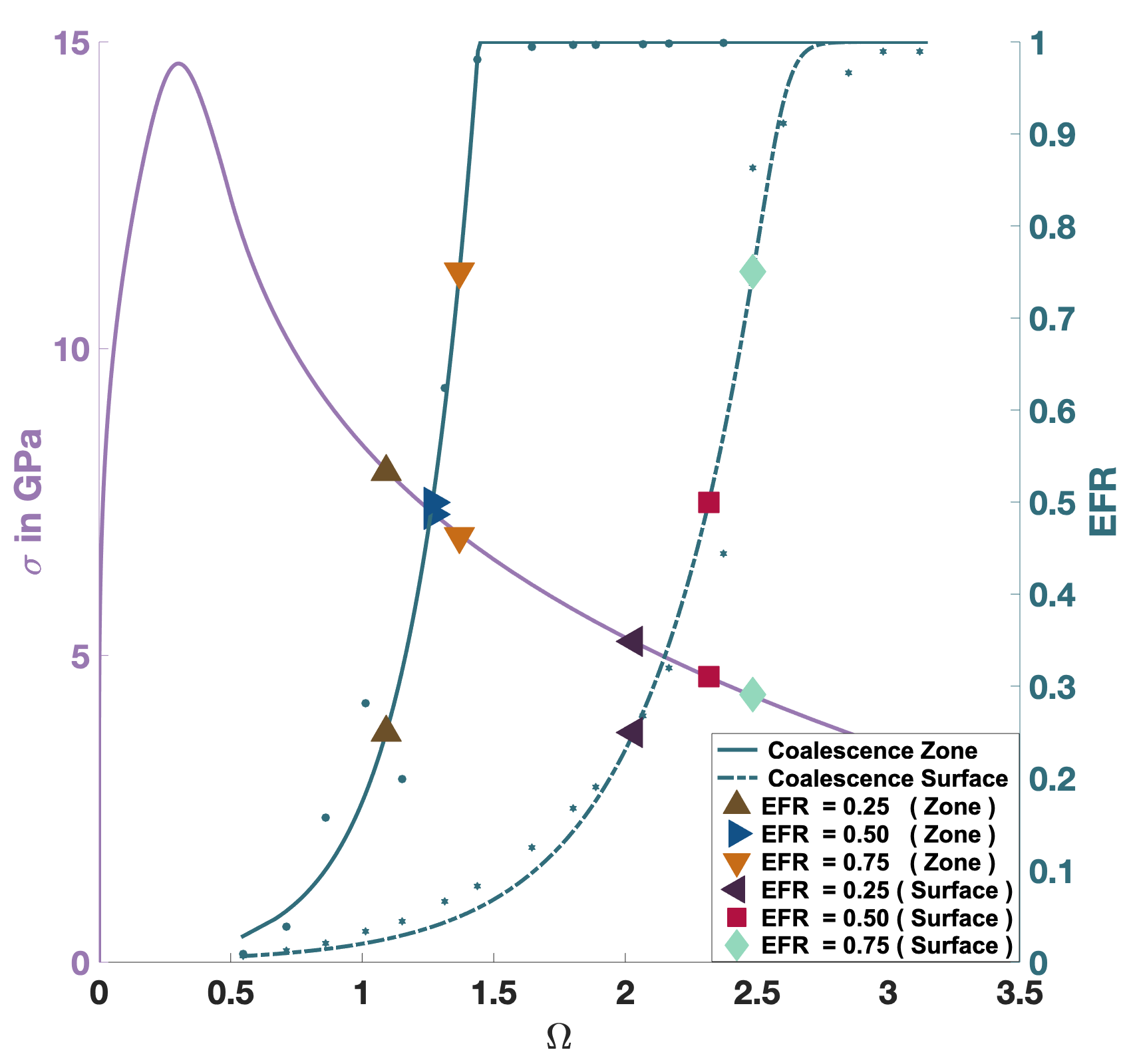}
 \caption{Variation of stress and EFR with damage. The solid and dotted lines in the EFR v/s damage plot correspond to Richard’s 5-parameter fit for coalescence zone and coalescence surface approaches respectively. The 0.25, 0.5 and 0.75 EFR states have been highlighted in the stress-damage and EFR-damage curves for both the coalescence zone and coalescence surface approaches.}
 \label{fig:S_E_D}
 \end{figure}

The mean fragment size ($d_{frag}$) at a given initial defect size is non-unique with damage (Fig: \ref{fig:SDS1}). This suggests that there is some other characteristic quantity that has a more unique relationship with fragment statistics than damage. One such possible measure is the degree of fragmentation, captured here by the effective fragmentation ratio (EFR). The EFR is defined as the ratio of the volume occupied by all but the largest fragment to the simulation box volume (Fig: \ref{fig:Efr}). The largest fragment, appearing in gray in Fig: \ref{fig:Efr}, provides insight into the onset of fragmentation. When very little fragmentation has occurred, the largest fragment occupies a large part of the volume, encloses other fragments and is connected to the boundary of the simulation box (Fig: \ref{fig:Efr_inc} a,b). When significant fragmentation has occurred, the largest fragment is not necessarily connected to the boundary (Fig: \ref{fig:Efr_inc} c). However, it is worth noting that a smaller simulation box might also lead to lower EFR values, as it may not be large enough to capture the tail end of the fragment size distribution. In these simulations, the simulation box is at least 10 times the length of the largest wing crack (not the average). Fig: \ref{fig:S_E_D} shows the evolution of stress with damage as well as the corresponding EFR values for both the coalescence zone approach and the coalescence surface approach. As expected, EFR increases with damage. The 5-parameter Richard’s asymmetric growth curve \shortcite{Richards1959} has been used to fit the EFR values (solid and dotted EFR lines in Fig: \ref{fig:S_E_D}):
\begin{equation}
  \label{eq:Richard5param}
EFR={\Xi}_0+\frac{{\Xi}_{\infty}}{\Big(1+\Omega_{\phi}e^{k_g(\Omega-\Omega_m)}\Big)^{1/\Omega_{\phi}}},
\end{equation}
\noindent
where ${\Xi}_0$ is the lower asymptote, ${\Xi}_{\infty}$ is the upper asymptote, $\Omega$ is the damage, $\Omega_m$ is the damage at maximum growth, $k_g$ is the growth rate and $\Omega_{\phi}$ is a variable which fixes the point of inflection. \par
\begin{figure}
\centering
 \includegraphics[width=0.75\textwidth]{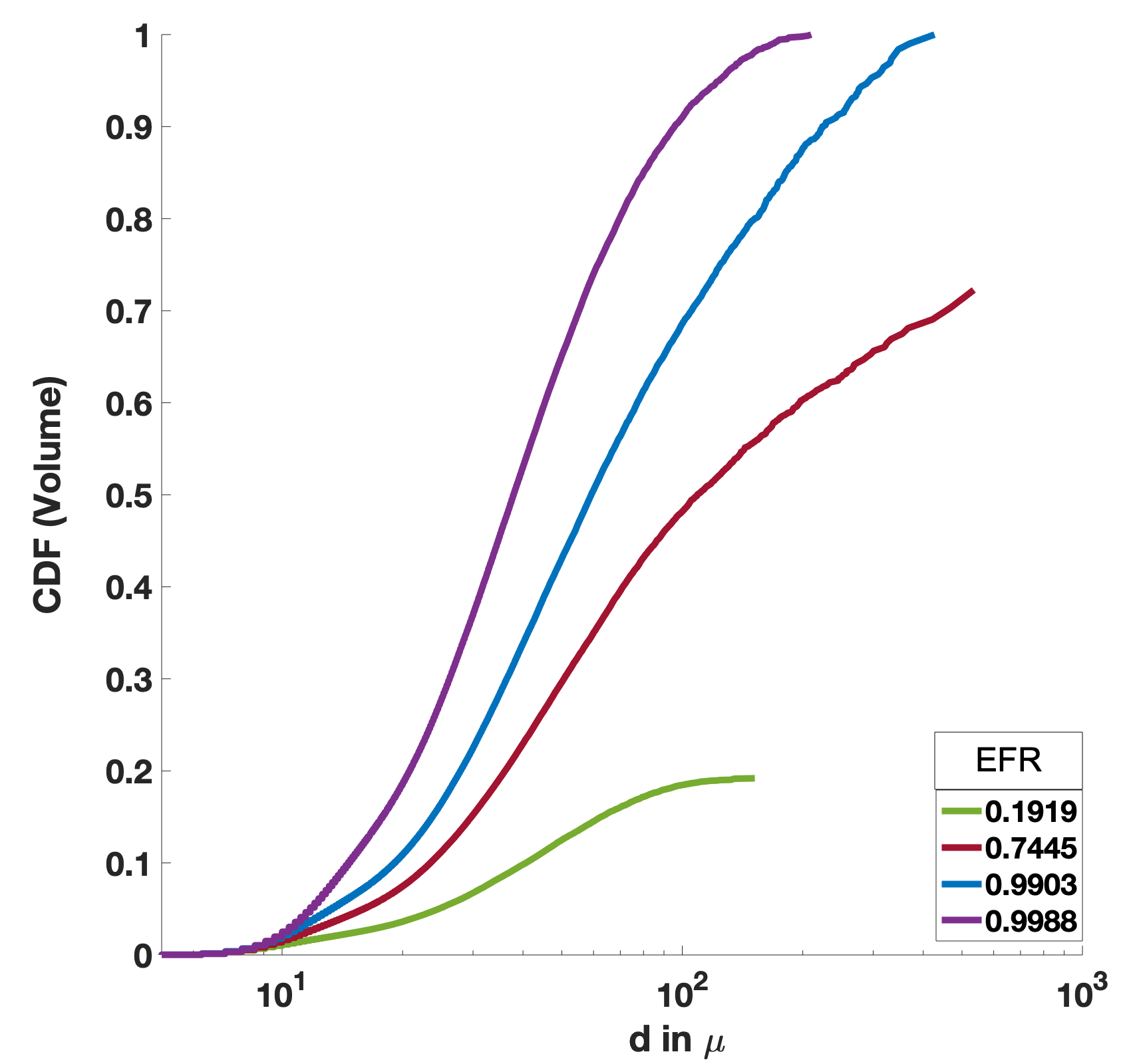}
 \caption{Volume CDF or fraction passing with fragment size, excluding the largest fragment, for different EFR values}
 \label{fig:Size_CDF}
 \end{figure}

For the current problem ${\Xi}_0=0$, ${\Xi}_{\infty}=1$, so there are only three parameters that require fitting. The fitted EFR-damage curve has been used to interpolate and obtain damage thresholds corresponding to any given EFR value. The 0.25, 0.5 and 0.75 EFR values have been highlighted along the EFR-damage curves as well as the stress-damage curve. It is observed that any significant fragmentation will mostly occur on the post peak strength part of the stress-strain curve. Fig: \ref{fig:SDS2} shows that the mean fragment size at a given initial defect size bears a unique relationship with EFR, for a constant defect density. In most cases, for the coalescence zone approach, the jump from around EFR=0.7 to EFR=0.9 happens almost instantaneously as seen in Fig: \ref{fig:S_E_D}, suggesting a threshold EFR of 0.75 could be used to mark a sharp transition to granular phase. Fig: \ref{fig:Size_CDF} shows a plot of the fraction passing by weight with fragment size ($d$) for different EFR values at a 3D crack density, $\eta=20x10^{12}\, cracks/m^3$ and strain rate of $10^6s^{-1}$. The curves are obtained from the discrete CDF of fragment volumes excluding the largest fragment. When the curves are smooth and do not have an outlier fragment, causing them to terminate far below one, it can be argued that all the fragments follow a smooth distribution and the material is completely fragmented. Otherwise it will suggest that there are only some small fragments contained within a mostly intact material. It is obvious from the figure that at EFR<0.75 the material hasn’t completely fragmented, whereas at EFR>0.9 it can be argued that the material has completely fragmented. EFR=0.75 is close to the highest EFR value at which an outlier fragment in the CDF of fragment volumes is clearly discernible. Any further fragmentation will be a consequence of particle breakage during granular flow where granular mechanics dominate. This further supports the idea of granular transition at an EFR of 0.75.

\subsection{Granular Phase Transition}

The fragmentation model contains a number of microstructural parameters, such as initial defect size ($2l_{initial}^{defect}$), initial three-dimensional crack density ($\eta$), polycrystalline fracture toughness ($K_{IC}$), strain rate ($\dot{\epsilon}$) and elastic modulus (E). Significant changes in the dependence of EFR on damage were only observed when changing  $l_{initial}^{defect}$ and $\eta$. Two-dimensional damage is calculated as,
\begin{equation} 
  \label{eq:Damage_2D_Dis}
\Omega=\displaystyle\sum_{j}\displaystyle\sum_{k}\eta(k,\theta_j)^{2/3}l_w(k,\theta_j)^2,
\end{equation}
where $\eta(k,\theta_j)$ and $l_w(k,\theta_j)$ are the 3-dimesnional crack density and the wing crack length corresponding to the $k_{th}$ initial defect size and $j_{th}$ initial defect orientation. $\eta^{2/3}$ represents the two-dimensional crack density analogue. 
Initial damage has been defined as
\begin{equation} 
  \label{eq:Om_i}
\Omega_i=\displaystyle\sum_{j}\displaystyle\sum_{k}\eta(k,\theta_j)^{2/3}l_{initial}^{defect}(k,\theta_j)^2.
\end{equation}
The representative initial crack length is defined as, 
\begin{equation} 
  \label{eq:Initial_l}
l_i=\sqrt{\frac{\Omega_i}{\eta^{2/3}}}.
\end{equation}
The representative final crack length is defined as, 
\begin{equation} 
  \label{eq:Final_l}
l_f=\sqrt{\frac{\Omega_f}{\eta^{2/3}}},
\end{equation}
where $\Omega_f$ is the $\Omega$ at transition.\par
$l_i$ and $l_f$ are the root mean square values of the initial and final crack lengths respectively.
Using these expressions, for random orientation of initial defects, after performing an exhaustive set of fragmentation simulations for various strain rates, initial defect populations, elastic moduli and fracture toughness, the following fit was found for granular transition at EFR = 0.75:
\begin{equation} 
  \label{eq:Trans_Dmg_Stress}
\Omega_i=293.6{\Omega_{f_{ 0.75}}}^{-1.049}{\frac{\sigma_{f_{ 0.75}}\eta^{-1/6}}{K_{IC}}}^{-2.784},
\end{equation}
where $\Omega_{f_{ 0.75}}$ and $\sigma_{f_{ 0.75}}$ represent the state variables for transition damage and transition stress at EFR = 0.75.\par
The adjusted R-squared value for this expression is 0.96. However this definition is difficult to apply in practice, since it requires knowledge of both the damage and stress at an EFR of 0.75. Alternatively, a simpler function in terms of the representative crack length and the initial defect distribution was used to provide an R-squared value of 0.92:
\begin{equation} 
  \label{eq:Trans_Dmg_Fit_Ran}
l_f=1.354\eta^{-1/3}-0.6977l_i .
\end{equation}
The  results have been compared against the case of fixed flaw orientation along the most favorable direction in \shortciteA{PALIWAL2008} using a similar parametric study of the fragmentation model. This leads to the transition fit with an adjusted R-squared value of 0.98:
\begin{equation} 
  \label{eq:Trans_Dmg_Fit_Del}
l_f=1.576\eta^{-1/3}-1.095l_i .
\end{equation}
Although, the trend appears to be similar to the random orientation case, the coefficients are slightly different. A more general form of the transition equation will be discussed in Sec: \ref{sec:Disc}.\par
In the following section, a phenomenological fragmentation model will be discussed. It predicts a similar form of the transition criterion as Eq:\ref{eq:Trans_Dmg_Fit_Ran}.

\subsection{Phenomenological model for transition}

We propose a simple phenomenological model for granular transition. The initial defect size is taken to be $2l_i$, the final crack length to be $2l_f$ and the three-dimensional crack density to be $\eta$. It is assumed that a given percentage of the defect centers need to be connected for fragmentation to occur. This assumption is arbitrary and can be set to any given value to meet a certain degree of fragmentation.
For random orientation, $\theta$, and initial defect size, $l_i$, the mean vertical projection of a defect is  
\begin{equation} 
  \label{eq:Project_li}
\frac{2}{\pi}\int_{0}^{\pi/2}l_i\cos{\theta} d\theta = 0.637l_i.
\end{equation}
Assuming defect locations are uncorrelated and follow a Poisson process, the defect spacing is an exponentially distributed random variable with mean spacing $\frac{1}{\lambda}=\eta^{-1/3}$,
\begin{equation} 
  \label{eq:exp_dist}
f_x(x)=\lambda e^{-\lambda x}.
\end{equation}
Therefore the cumulative density function is 
\begin{equation} 
  \label{eq:exp_cum_dist}
F_x(x)=1-e^{-\lambda x}.
\end{equation}
The spacing that corresponds to 90\% crack coalescence is $x=-\frac{ln(0.25)}{\lambda}=2.3026\eta^{-1/3}$.\par
In other words, for fragmentation to occur, 90\% of all defect center spacings have to be connected to each other. In order for that to happen, the final crack length plus the mean vertical projection of the initial defect size should therefore become equal to $2.3026\eta^{-1/3}$. Hence,\\
$2l_f+2*0.637l_i=2.3026\eta^{-1/3}$.\\
So, the material transitions when
\begin{equation} 
  \label{eq:PhenTrans90}
l_f=1.151\eta^{-1/3}-0.637l_i.
\end{equation}
If we assume that the criterion for fragmentation involves 75\% of the the defect centers to be connected to each other, the transition criterion can be rewritten as
\begin{equation} 
  \label{eq:PhenTrans75}
l_f=0.693\eta^{-1/3}-0.637l_i.
\end{equation}
These expressions have a similar form to Eqs:\ref{eq:Trans_Dmg_Fit_Ran} and \ref{eq:Trans_Dmg_Fit_Del} , though the coefficients are underestimated.\par
\begin{figure}
\centering
 \includegraphics[width=0.5\textwidth]{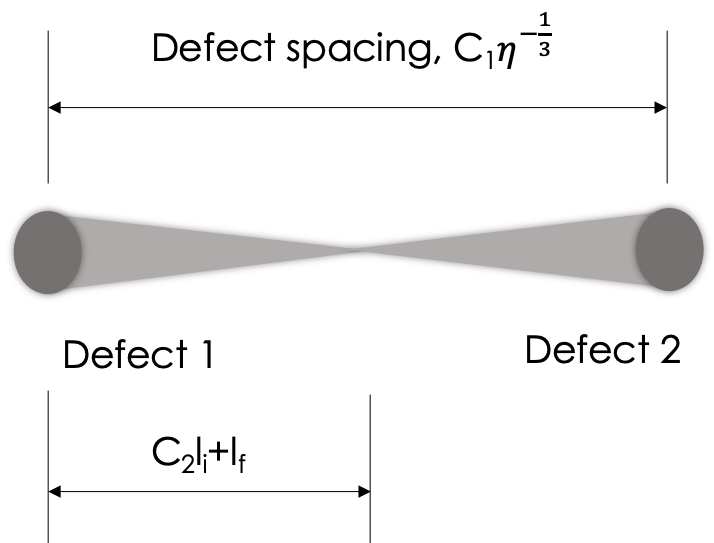}
 \caption{Schematic representation of phenomenological granular transition model}
 \label{fig:Phen_Model}
 \end{figure}

In general, the fragmentation criterion can be expressed as $2(l_f+C_2 l_i)=C_1 \eta^{-1/3}$.  Fig: \ref{fig:Phen_Model} shows a schematic representation of the problem.
The constants $C_1$ and $C_2$ are determined by the percentage of cracks that need to be connected in order to achieve a certain degree of fragmentation and the percentage of initial defects that actually get activated. The latter would depend on the flaw friction, confining stresses and also to some extent on the strain rate. The constant $C_2$ will depend on the orientation of initial defects. It is close to 0.637 for random defect orientation and $\cos{(\theta_{fixed})}$ for a fixed defect orientation, $\theta_{fixed}$.  

\section{Results - Fragment statistics}
\begin{figure}
\centering
 \includegraphics[width=0.75\textwidth]{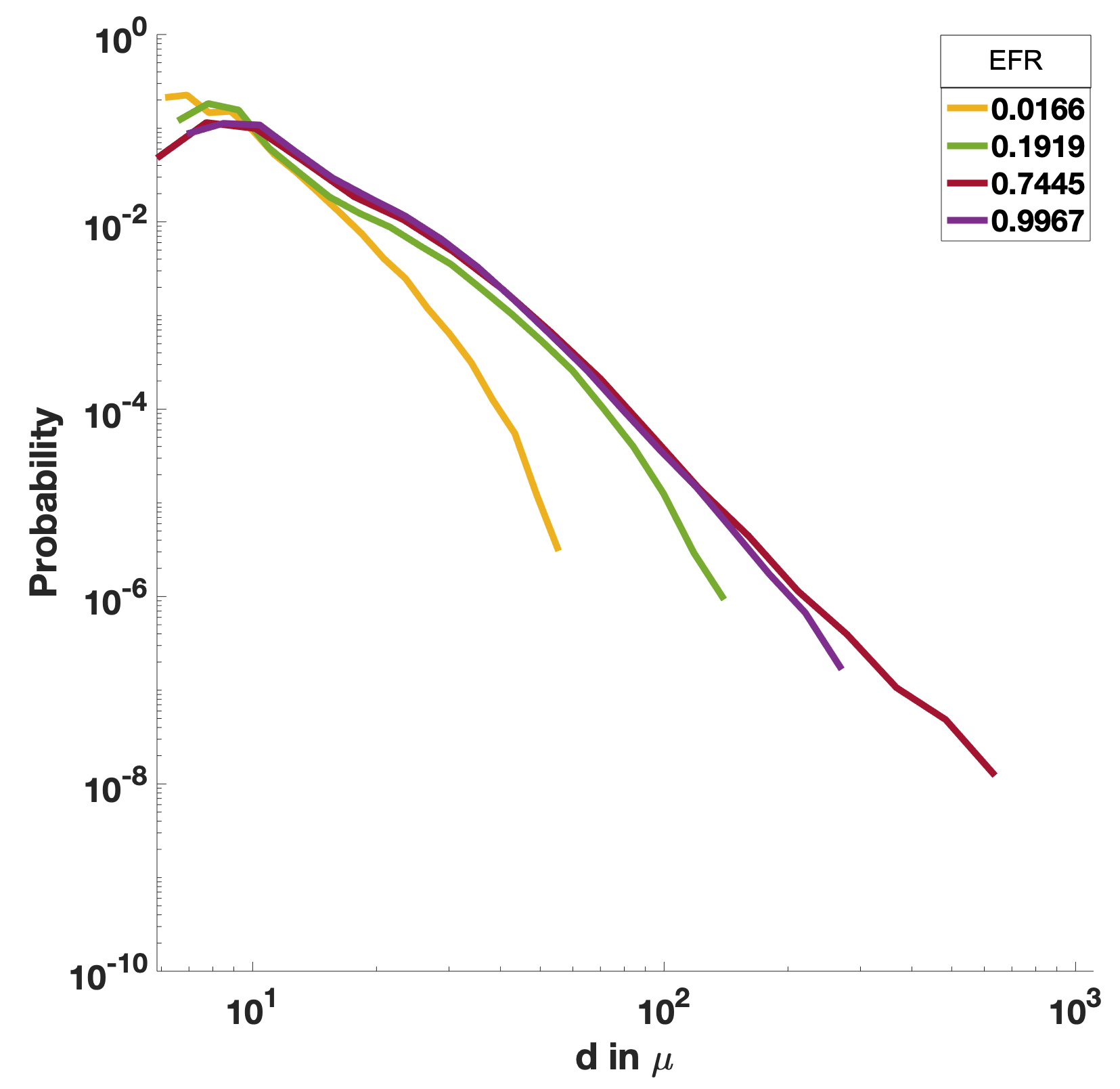}
 \caption{Fragment size distribution at $\eta=20x10^{12}\, cracks/m^3$ and $\dot{\epsilon}=10^6 s^{-1}$ for varying EFR }
 \label{fig:Frag_Size_Prob}
 \end{figure}

In the numerical fragmentation model, the connected regions correspond to individual fragments. The total number of voxels in a bounded region times the volume of each voxel is a measure of fragment volume. Fragment size has been computed as the cube root of fragment volume. Fragment size distribution, obtained for different values of initial defect sizes and strain rates, always followed a power law distribution, except for the smallest fragment sizes, which likely reflects the limitation of the selected resolution size. Fig: \ref{fig:Frag_Size_Prob} shows the fragment size distribution at different EFR values for 3D crack density, $\eta=20x10^{12}\, cracks/m^3$ at a strain rate of $10^6s^{-1}$. The largest fragment has been omitted for these plots. 
As mentioned before, the mean fragment size ($d_{frag}$) at a given EFR and flaw density does not seem to depend significantly on the initial defect size (Fig: \ref{fig:SizeDefectSize}). This should not be confused with the final fragment size distribution obtained from experiments with different initial defect sizes. This is because post granular phase energy dissipation due to refragmentation might still be different, due to different residual energies in either case. However, there is a dependence of the mean fragment size ($d_{frag}$) v/s EFR with crack density values ($\eta$). This is not unexpected; for a similarly scaled system one might expect the mean fragment size ($d_{frag}$)  to scale with $\eta^{-1/3}$.  In reality a proportionate scaling of the system does not necessarily scale the local stress states similarly, and thus affects the coalescence zone. \par
\begin{figure}[htpb]
  \centering
    \subfloat[Variation with $\Omega$]{\label{fig:SCD1}\includegraphics[width=0.5\textwidth]{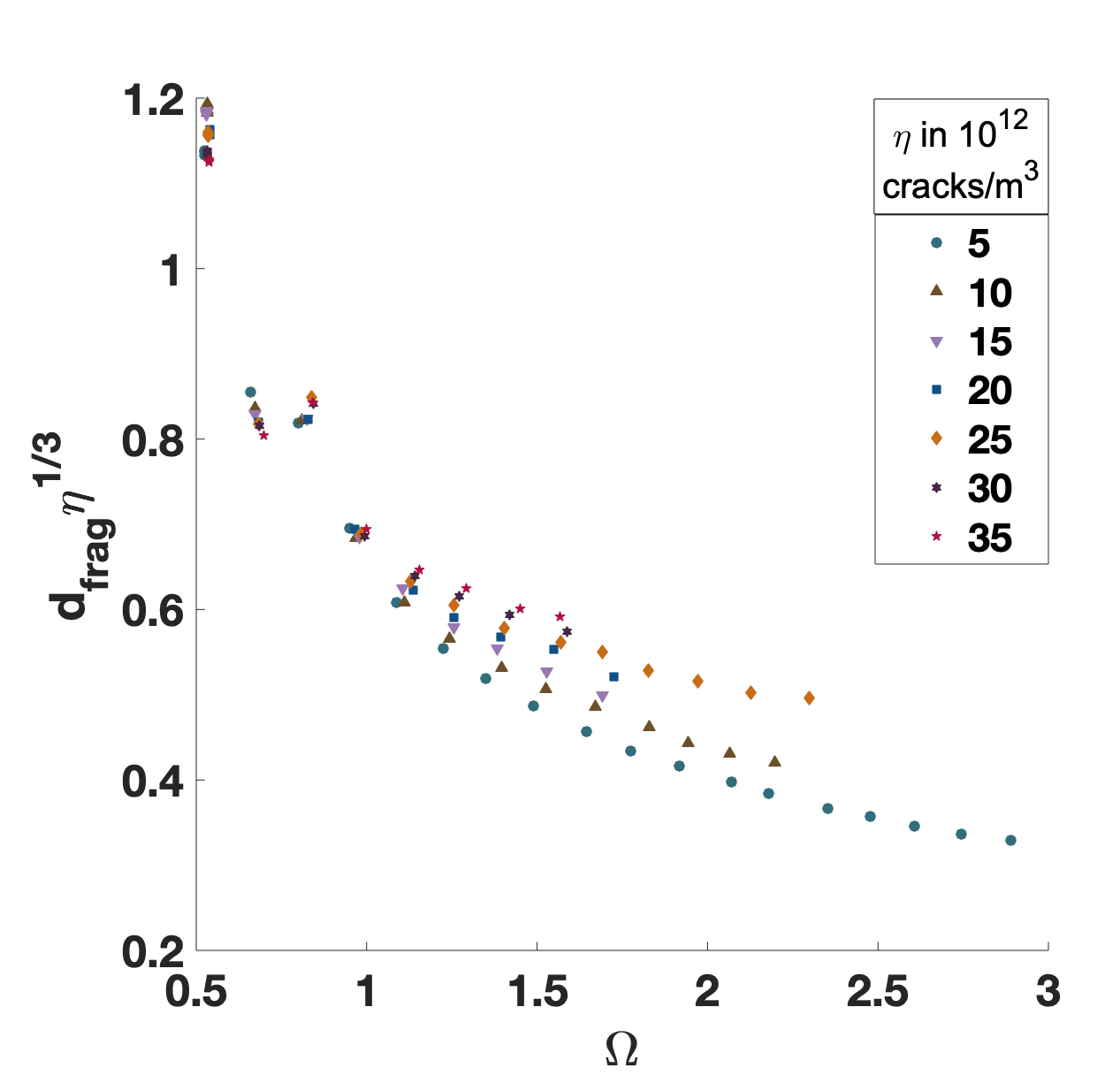}}\hfill
    \subfloat[Variation with EFR]{\label{fig:SCD2}\includegraphics[width=0.5\textwidth]{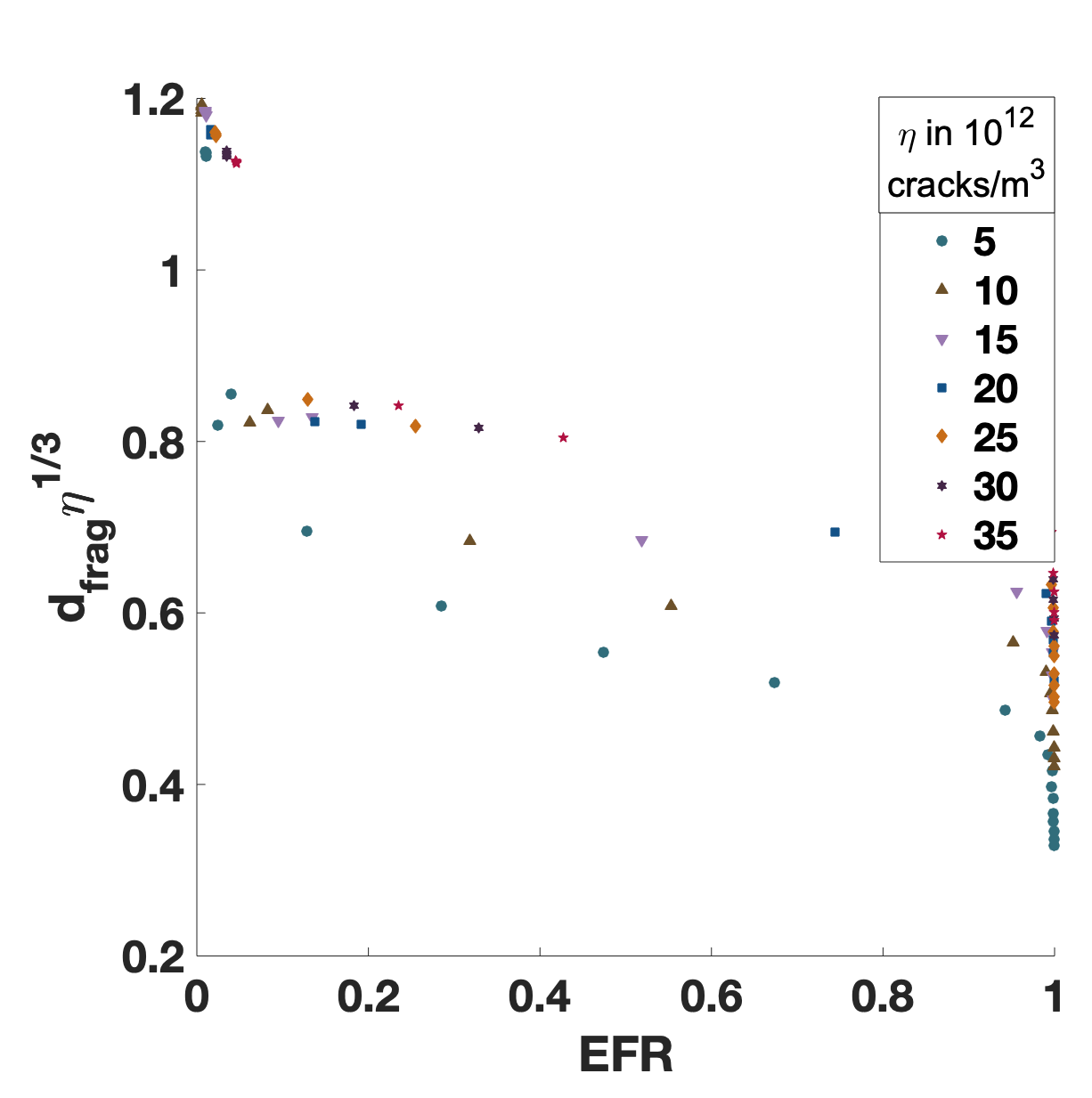}}\hfill 
  \caption{Variation of scaled mean fragment size with damage and EFR at initial defect size of $10\mu$ and $\dot{\epsilon}=10^6 s^{-1}$, for different crack densities ($\eta$)}
  \label{fig:SizeCrackDensity}
\end{figure}
\begin{figure}[htpb]
  \centering
    \subfloat[Variation with $\Omega$]{\label{fig:MDS1}\includegraphics[width=0.5\textwidth]{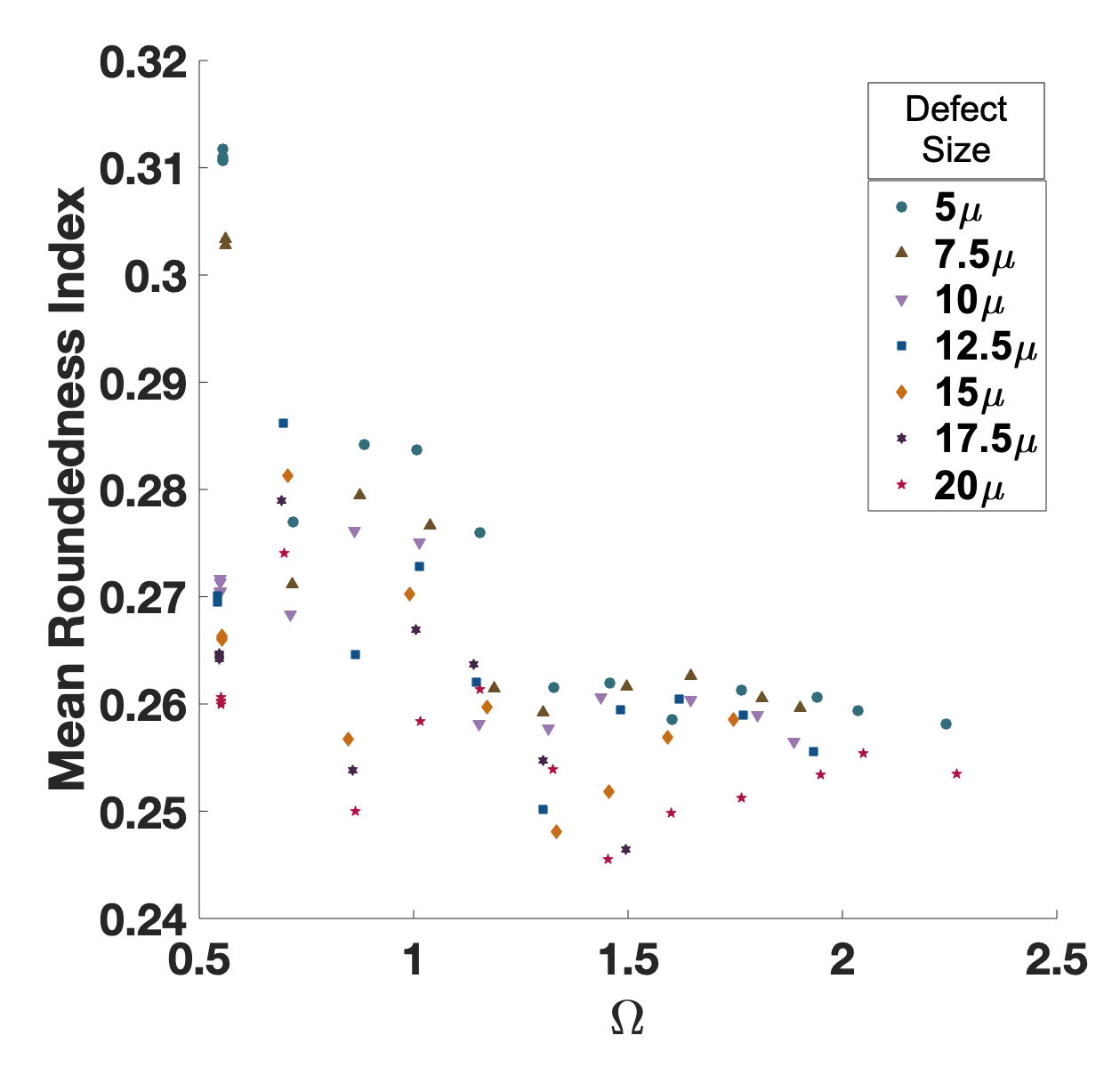}}\hfill
    \subfloat[Variation with EFR]{\label{fig:MDS2}\includegraphics[width=0.5\textwidth]{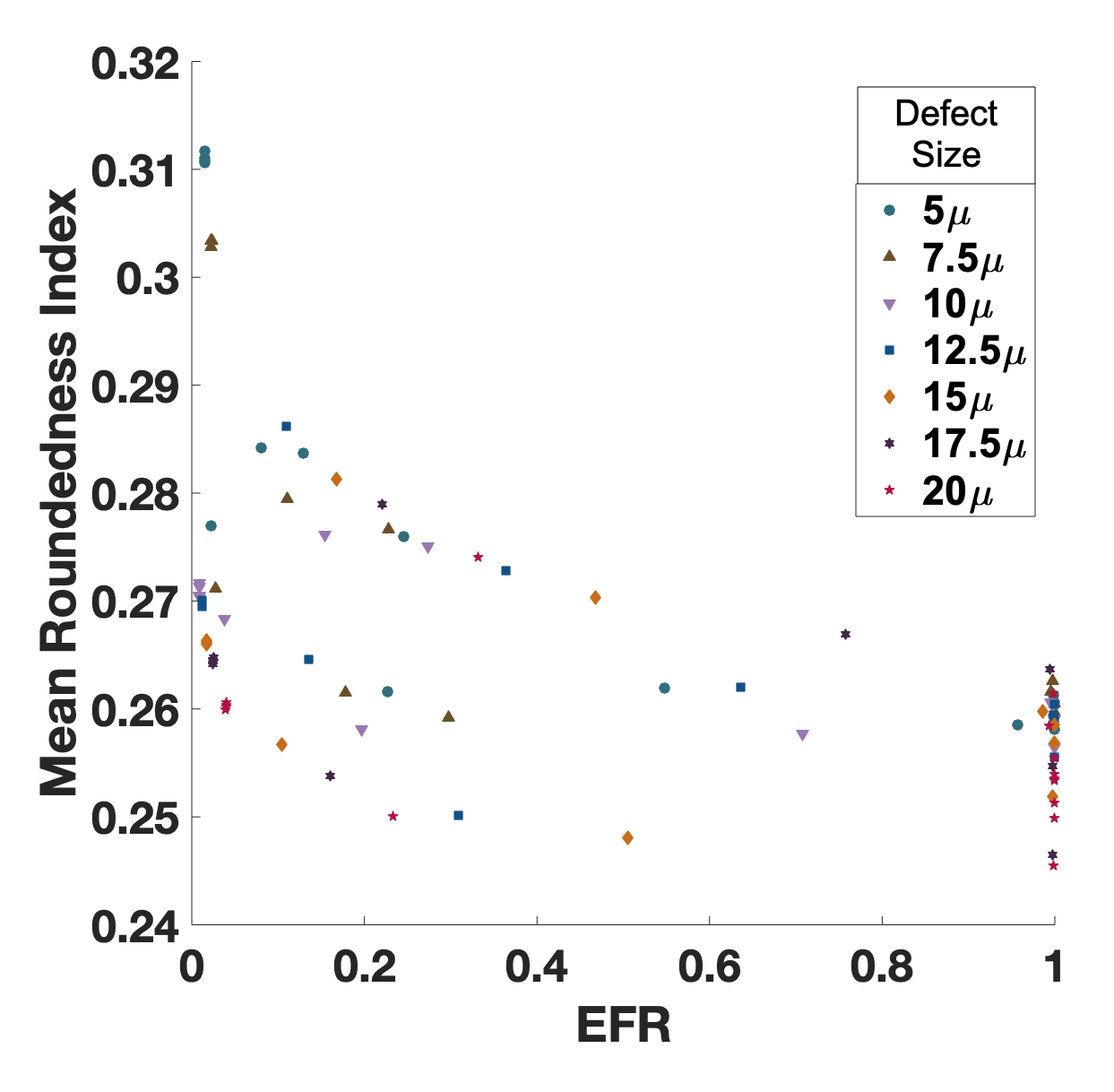}}\hfill 
  \caption{Variation of mean roundedness index with damage and EFR at  $\eta=22x10^{12}\, cracks/m^3$ and $\dot{\epsilon}=10^6 s^{-1}$ for different initial defect sizes}
  \label{fig:MRIDefectSize}
\end{figure}
\begin{figure}[htpb]
  \centering
    \subfloat[Variation with $\Omega$]{\label{fig:SolDS1}\includegraphics[width=0.5\textwidth]{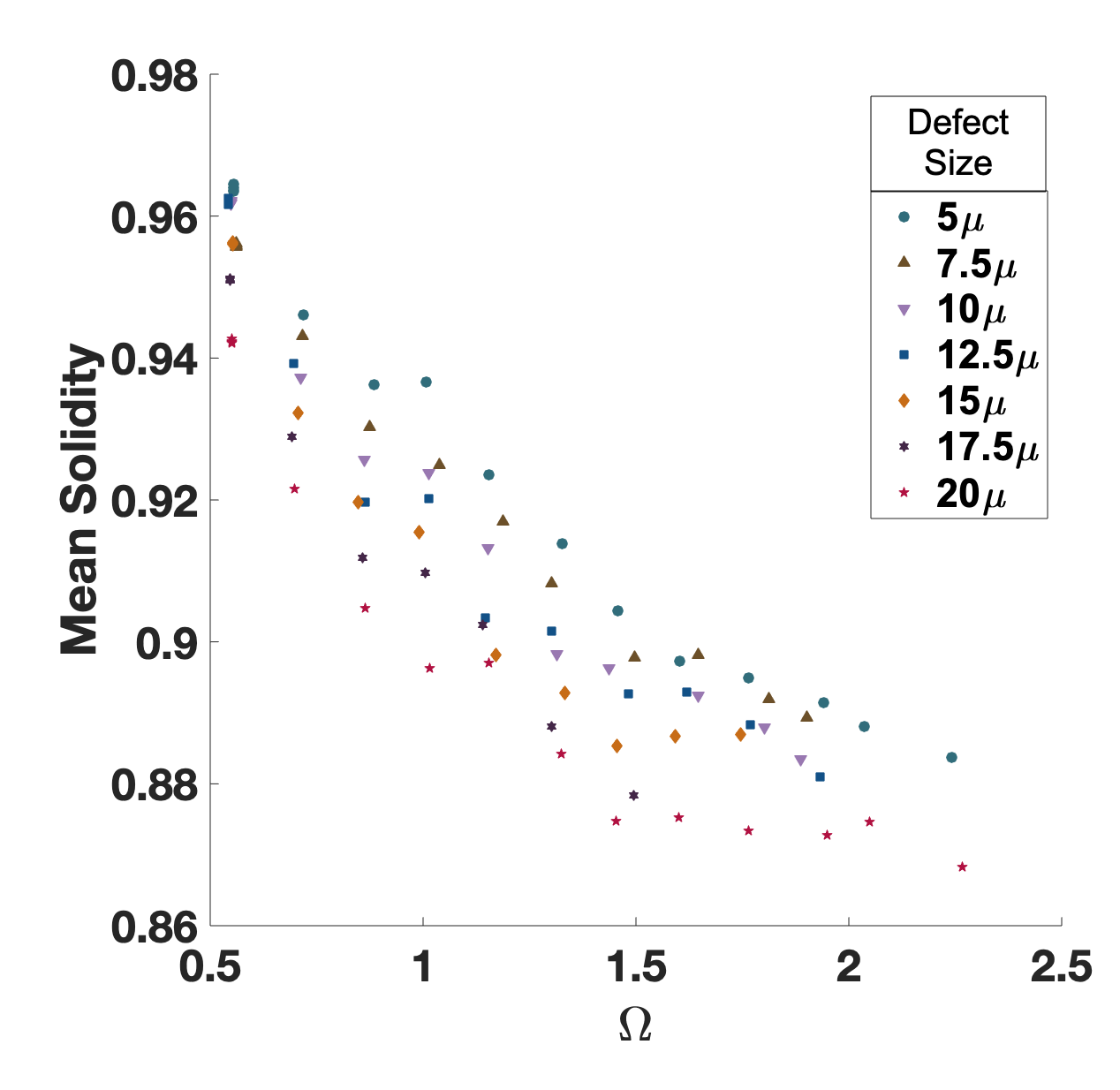}}\hfill
    \subfloat[Variation with EFR]{\label{fig:SolDS2}\includegraphics[width=0.5\textwidth]{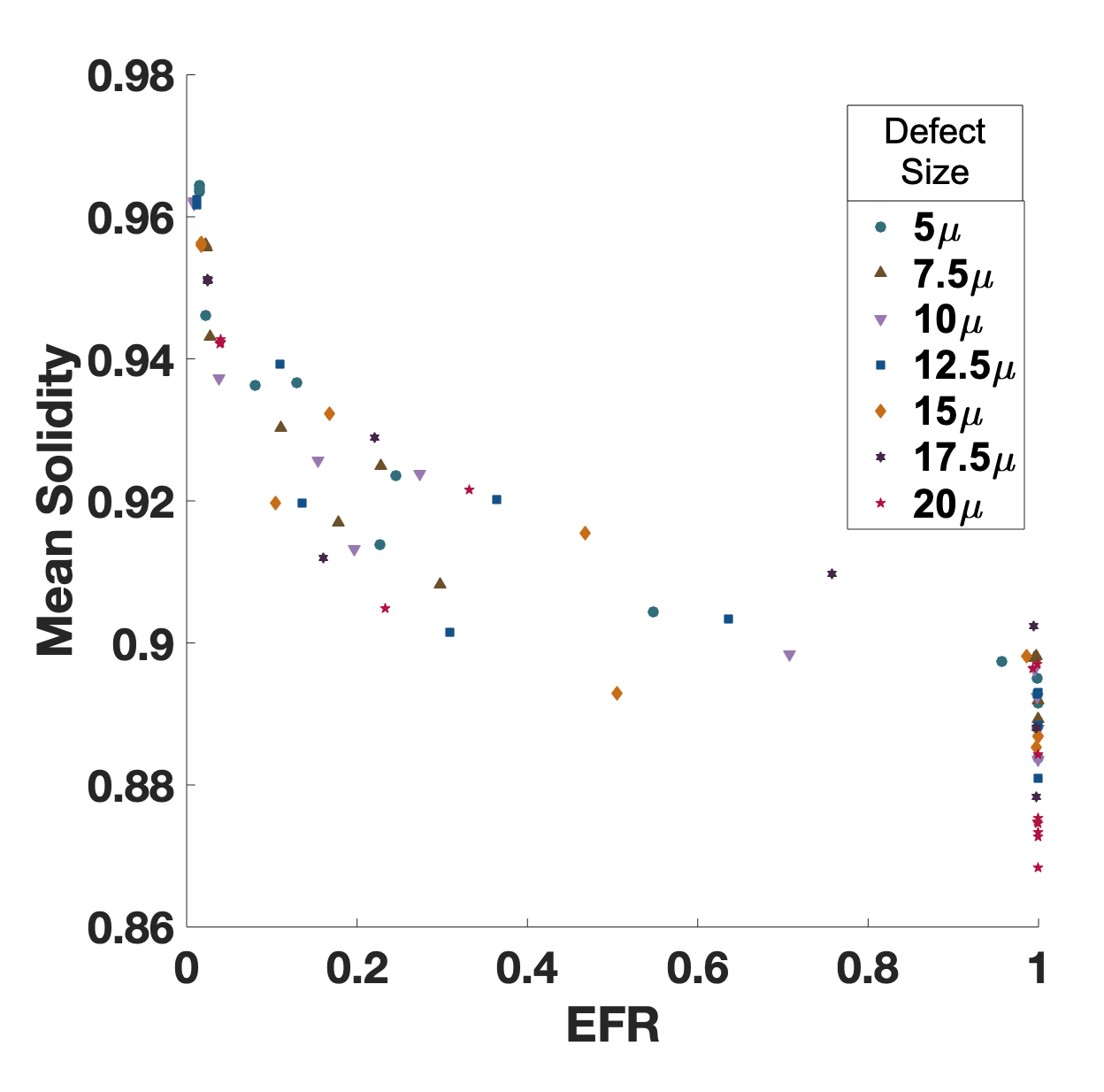}}\hfill 
  \caption{Variation of mean solidity with damage and EFR at  $\eta=22x10^{12}\, cracks/m^3$ and $\dot{\epsilon}=10^6 s^{-1}$ for different initial defect sizes}
  \label{fig:SolDefectSize}
\end{figure}
\begin{figure}[htpb]
  \centering
    \subfloat[Variation with $\Omega$]{\label{fig:MCD1}\includegraphics[width=0.5\textwidth]{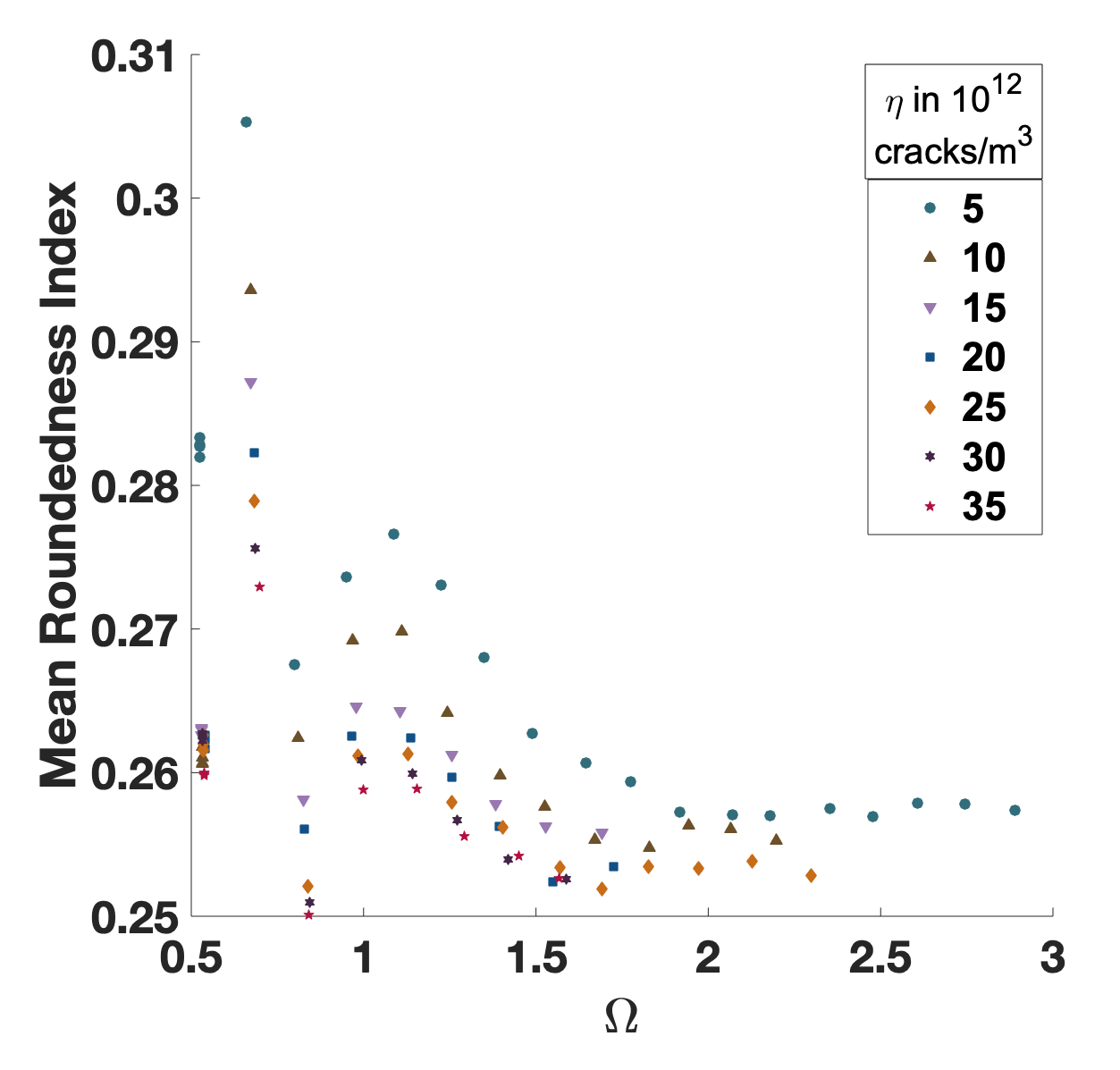}}\hfill
    \subfloat[Variation with EFR]{\label{fig:MCD2}\includegraphics[width=0.5\textwidth]{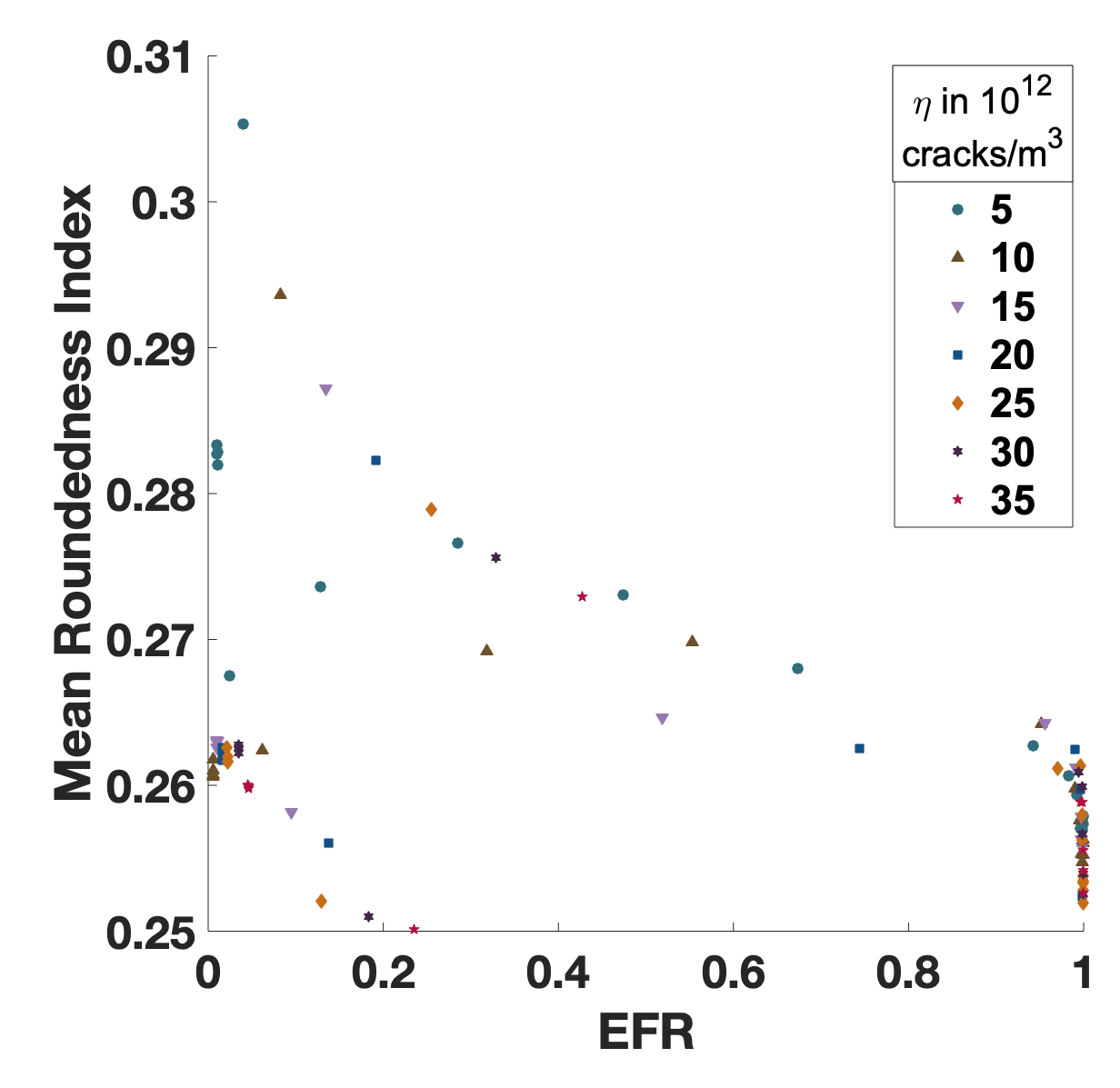}}\hfill 
  \caption{Variation of mean roundedness index with damage and EFR at initial defect size of $10\mu$ and $\dot{\epsilon}=10^6 s^{-1}$ for different crack densities ($\eta$)}
  \label{fig:MRICrackDensity}
\end{figure}
\begin{figure}[htpb]
  \centering
    \subfloat[Variation with $\Omega$]{\label{fig:SolCD1}\includegraphics[width=0.5\textwidth]{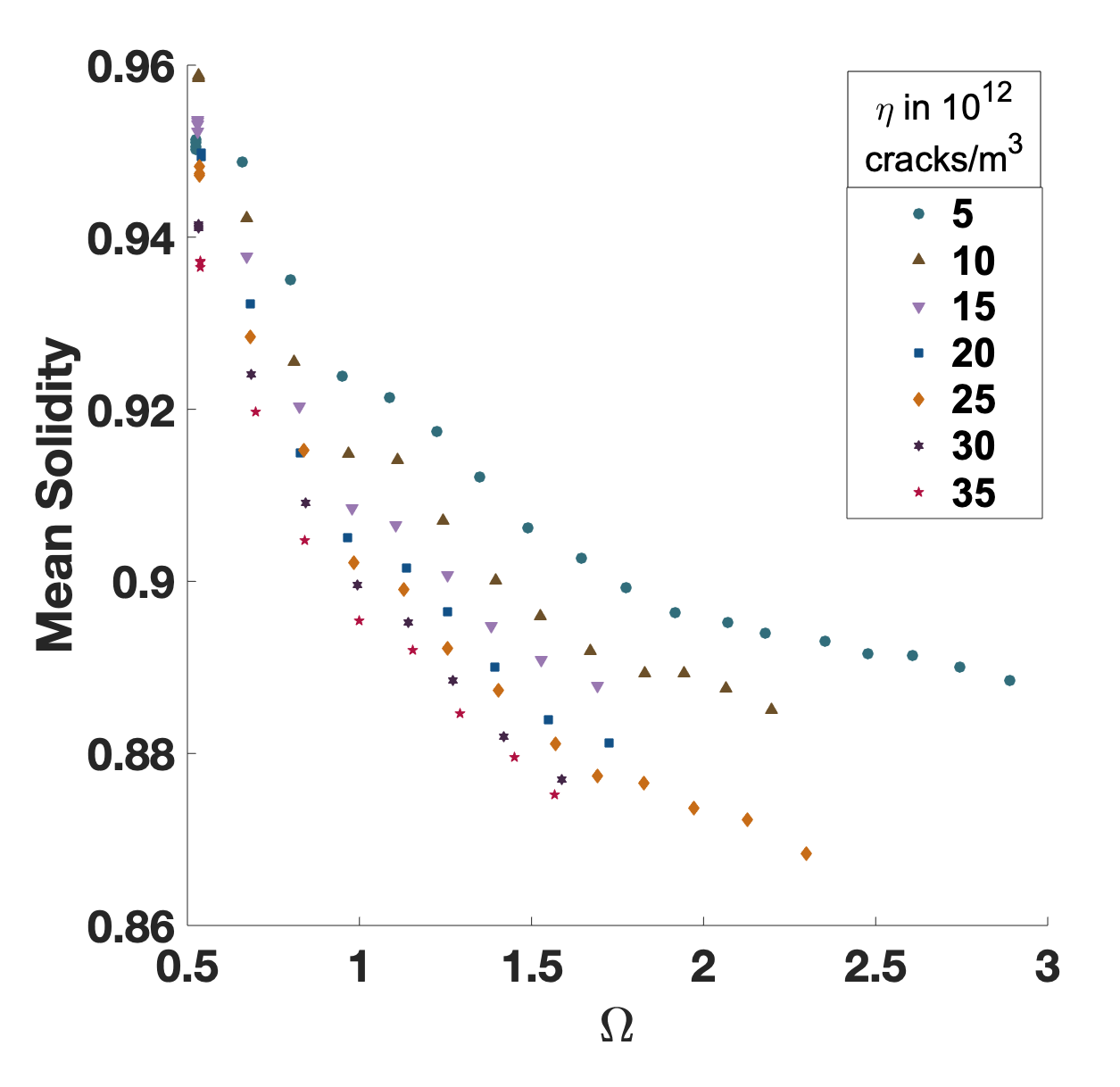}}\hfill
    \subfloat[Variation with EFR]{\label{fig:SolCD2}\includegraphics[width=0.5\textwidth]{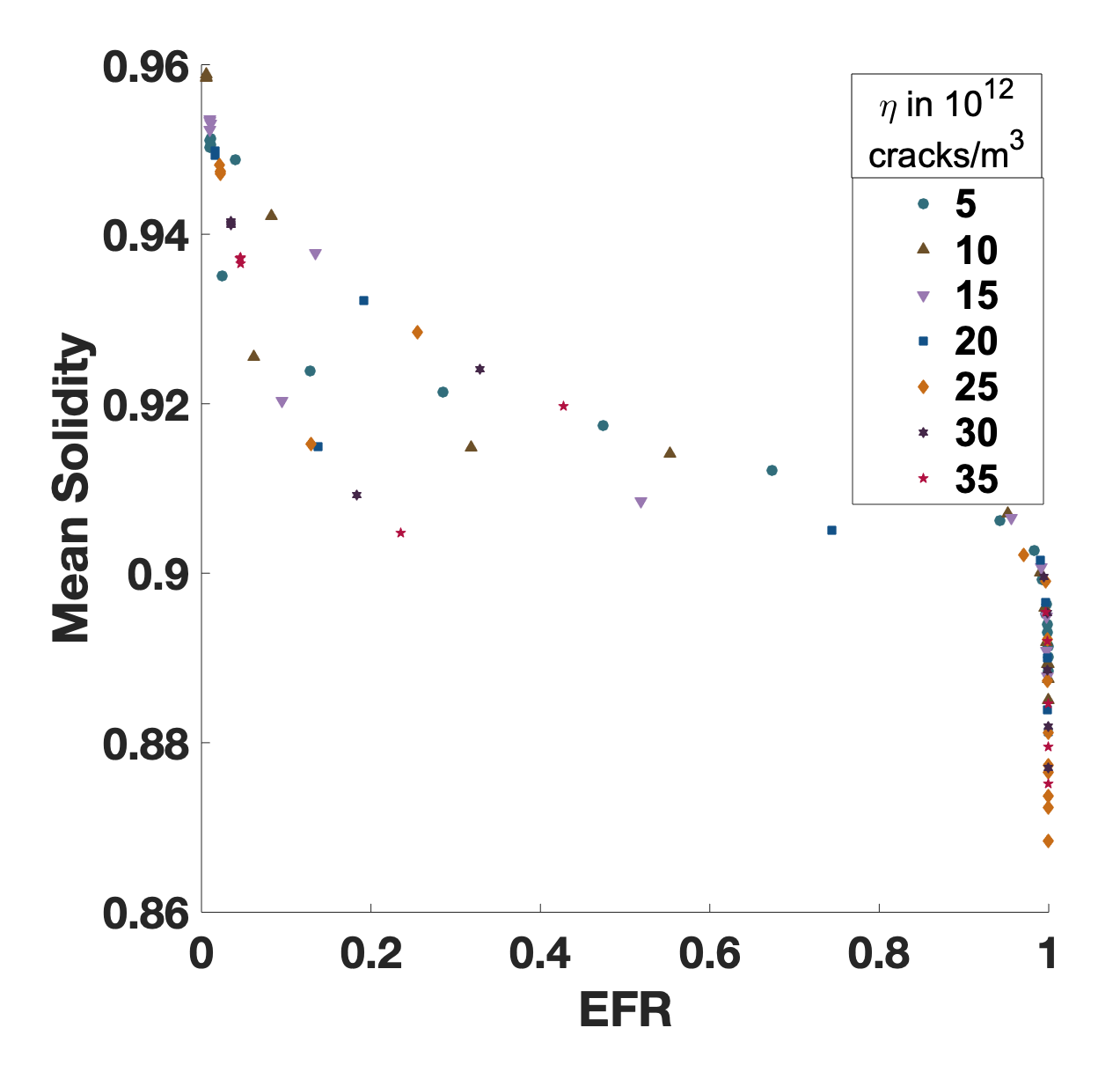}}\hfill 
  \caption{Variation of mean solidity with damage and EFR at initial defect size of $10\mu$ and $\dot{\epsilon}=10^6 s^{-1}$ for different crack densities ($\eta$)}
  \label{fig:SolCrackDensity}
\end{figure}

Fig: \ref{fig:SizeCrackDensity} shows the variation of scaled mean fragment size ( $d_{frag}\eta^{1/3}$)  with damage and EFR for different crack densities. It is worth noting that the initial defect size was not scaled proportionately, and the same value was used.
Other fragment properties like solidity, mean roundedness index can also be extracted. These properties have a more complicated relationship with EFR for different initial defect sizes and defect densities. Figs: \ref{fig:MRIDefectSize}, \ref{fig:SolDefectSize} show the variation of mean roundedness index and solidity with EFR and damage for different initial defect sizes (or half flaw size) respectively. Figs: \ref{fig:MRICrackDensity}, \ref{fig:SolCrackDensity} show the same for crack density. The general trend appears to suggest that roundedness and solidity decrease with increase in damage (or EFR) for given value of crack density and initial defect size except for some minor aberrations mostly observed at lower EFR values. This suggests that particles tend to be more angular at a higher level of fragmentation. Based on a transition EFR = 0.75 to 0.9, a mean solidity value of 0.91 and a mean roundedness index of 0.26-0.27 is suggested.

\subsection{Comparison with experiments} \label{ssec:CompExp}

\shortciteA{HOGAN2016} looks at the rate dependent fragmentation of boron carbide using a Split-Hopkinson pressure bar setup for confined and uniaxial compressive loading. The fragment size distribution for dynamic uniaxial compression best simulates the loading conditions in the current work. There are two distinct fragmentation regimes - a regime dominated by the processing induced microstructural flaws (Regime I) and another one that is dominated by boundary conditions and macroscopic structural failure (Regime II). The current work assumes periodic boundary conditions and is unable to account for problem specific macroscopic structural failure. The fragment size distribution generated from the current work is induced by the microstructural defect population modelled as micro-cracks and it is best compared with the Regime I fragments obtained from experiments. It has been assumed that the stress state in the experiments is homogeneous and simulates more or less a uniaxial loading condition. Also, any further fragmentation due to granular flow has been neglected or atleast assumed to not significantly alter the nature of the initial distribution of fragment sizes.

\begin{figure}
\centering
 \includegraphics[width=0.75\textwidth]{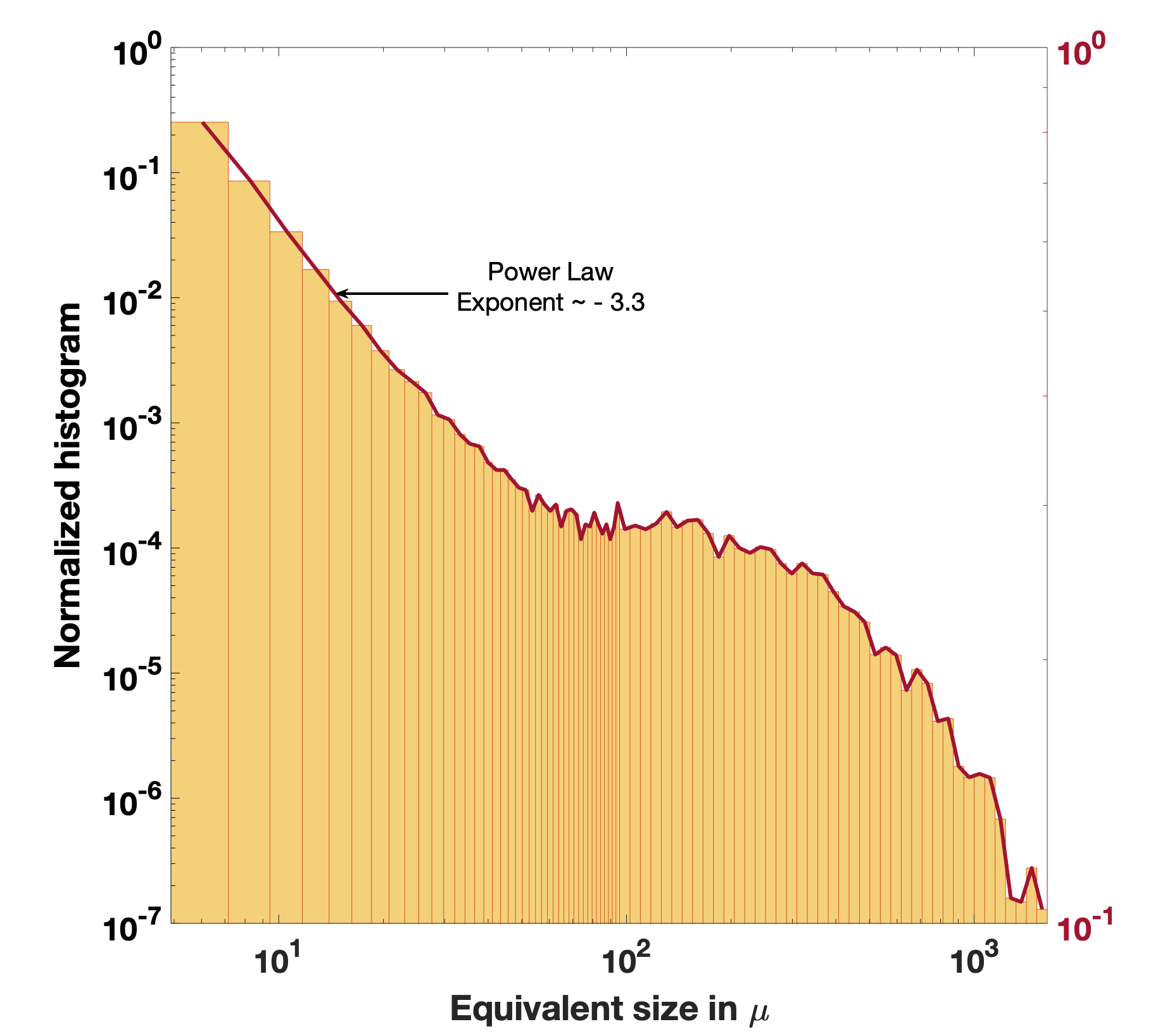}
 \caption{Fragment size distribution obtained from Kolsky bar dynamic fragmentation of Boron Carbide (Hogan et al., 2016)}
 \label{fig:KolskyBar_Hogan}
 \end{figure} 
 
 \begin{figure}
\centering
 \includegraphics[width=0.75\textwidth]{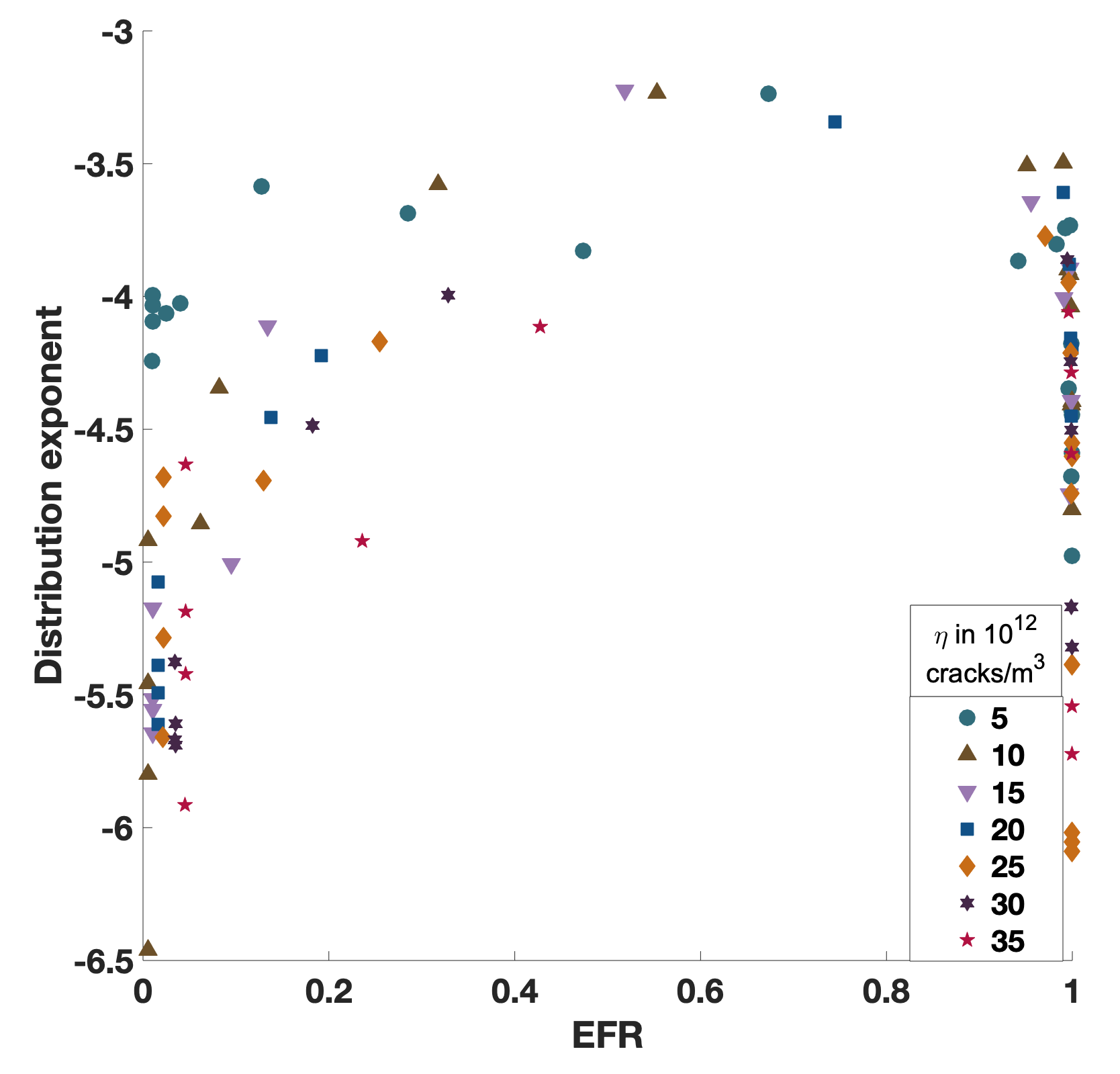}
 \caption{Power law distribution exponent of fragment sizes from numerical model}
 \label{fig:Dist_Exp}
 \end{figure} 
 
 \begin{figure}
\centering
 \includegraphics[width=0.75\textwidth]{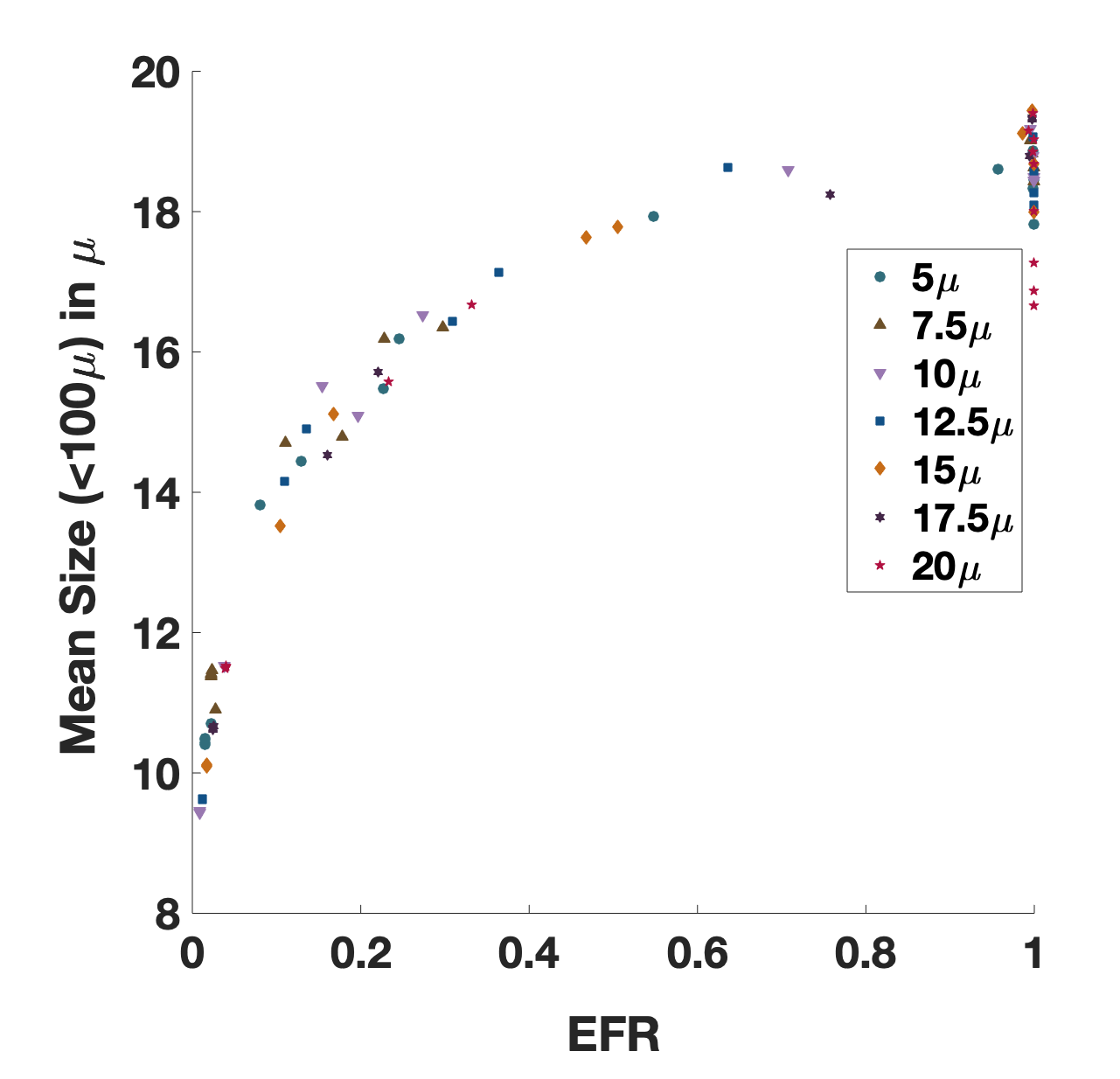}
 \caption{Mean volume averaged fragment size with change in EFR for fragments less than $100\mu$}
 \label{fig:Size_100}
 \end{figure}
\noindent
Fig: \ref{fig:KolskyBar_Hogan} shows the normalized histogram of fragment size distribution for dynamic uniaxial loading \shortcite{HOGAN2016}. The two distinct fragmentation zones can be easily demarcated. Fragments less than $100\mu$ (Regime I) appear to exhibit a power law relationship similar to our observations. The power law exponent for Regime I is around 3.3, whereas in the numerical model the exponent for EFR = 0.75 is around 3.5 (Fig: \ref{fig:Dist_Exp}). The volume-averaged mean size of fragments less than 100 $\mu$, computed from Hogan’s data is around 17.8 $\mu$. This agrees well with the corresponding mean size at around EFR=0.75 from the numerical model (Fig: \ref{fig:Size_100}). In Fig: \ref{fig:Size_100} the fragment size seems to initially increase with EFR, unlike in Fig:\ref{fig:SizeDefectSize}, because the fragments greater than $100\mu$ have not been included. As EFR increases, more of the larger fragments re-fragment to create smaller fragments. It is worth noting that the size reported for Hogan’s data corresponds to an equivalent diameter, while the size calculated from the numerical model is simply the cube root of the fragment volume. The numerical model is three dimensional while the fragment sizes calculated from Hogan’s experiments are obtained by calculating the major and minor axis lengths from two-dimensional projection of three-dimensional particles.

\section{Discussion} \label{sec:Disc}

The current approach treats granular transition as a near instantaneous mechanism. While it is not unreasonable to think that the mobilization of a few fragments accelerates fragmentation and subsequent granular behaviour, the process of crack coalescence might initiate much earlier. The model used to simulate dynamic crack growth ignores the modification of the number density of crack populations and their crack lengths. It remains unclear how sensitive the transition is to early stage coalescence. Early stage crack coalescence increases crack length but reduces crack density; these competing effects on damage might balance one another. However the competing effects might change the mode of crack propagation to one in which a wing crack growth-based damage model might be unable to account for. These are general limitations of wing-crack growth-based models. \par
\begin{figure}
\centering
 \includegraphics[width=0.75\textwidth]{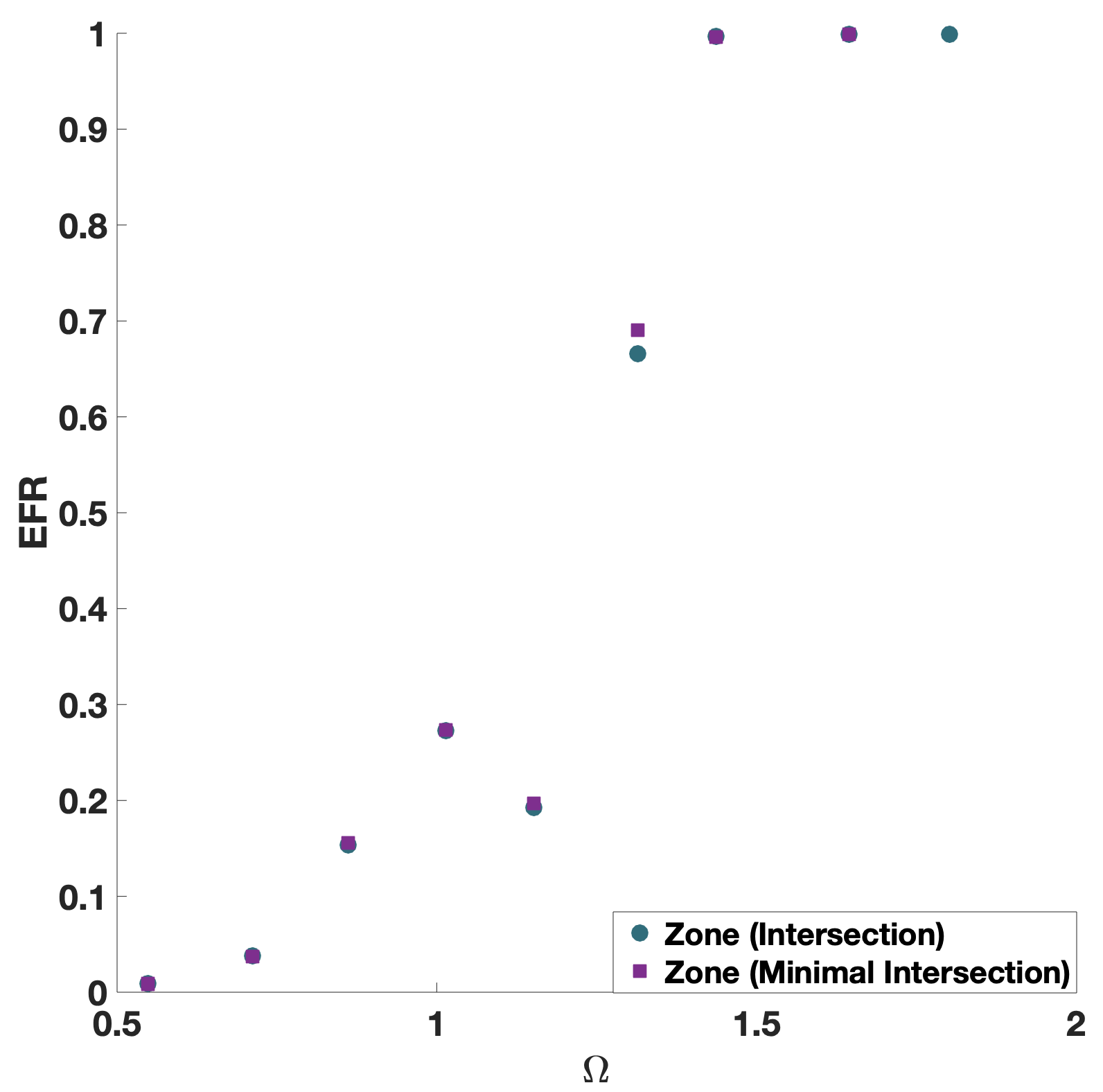}
 \caption{Comparison of simulating cracks with and without minimal intersection}
 \label{fig:Intersect}
 \end{figure}

The presented approach is generic and can also be exercised with other damage models by modifying the input crack statistics. In the numerical model, cracks have been generated randomly without any restriction on crack intersection. Simulating minimally intersecting cracks involves ensuring that most of the ellipses do not intersect with one another. This can be computationally challenging, especially at high crack density and for large crack lengths. This has been attempted in the current work by trying to generate cracks of a given size in the simulation box volume iteratively until there is no intersection with previously generated cracks. At every iteration the location is reset and at every ten iterations the $Q_x$ and $Q_y$ matrices are reset, essentially changing the crack orientation. If we fail to generate any such cracks in 500 iterations, the crack corresponding to the minimum number of intersecting voxels has been accepted. It is observed that generating minimally intersecting cracks does not affect the evolution of EFR with damage (Fig: \ref{fig:Intersect}). Given the computational challenge and the insensitivity of EFR to the constraint of minimal crack intersection, it makes sense to ignore the effect for parametric evaluation of fragmentation.\par
It has been observed that for the coalescence zone approach there are some outlier points in the EFR v/s damage curves (Fig: \ref{fig:CompApp}), where the EFR seems to drop before increasing again and following the general trend. EFR is not strictly monotonic with damage and the lack of monotonicity is not due to randomness in the model as this drop often happens around the same damage value for two independent set of simulations (Coalescence zone approach for different simulation box sizes in Fig: \ref{fig:CompApp}, different crack generation techniques in Fig: \ref{fig:Intersect}). A closer look at the coalescence zone approach suggests that this drop might be an artifact of the coalescence zones. While the coalescence zone approach is a much more efficient way of numerically accounting for crack coalescence, it treats the coalescence cracks as larger volumes. This is not a significant problem in general, but often the voxels associated with coalescence zones erode some of the smaller fragments. There are certain damage values around which the contribution of these zones towards creating new connections is overshadowed by them eroding the smaller fragments and later on reassigning those voxels to part of the largest fragment. It has been checked that when a sudden drop in EFR occurs, the number of fragments generated also decrease, which supports the above reasoning. Since the overall trend still remains monotonic, and the drop occurs at lower EFR values than those associated with fragmentation, the computational advantage of the coalescence zone approach outweighs the lack of monotonicity.

\par
Given the range of strain rates studied ($10^4$ to $10^6 s^{-1}$), significant strain rate dependency on the fragmentation criterion or fragment sizes at the onset of granular flow was not observed. This does not imply that strain rate does not affect fragmentation. Strain rate affects the stress-strain response, as well the residual stresses immediately after the onset of granular mechanics. This will influence further granular behaviour and/or further fragment breakage. However, strain rate does not seem to affect the transition damage values under high rate conditions. The ranges of strain rate studied ($10^4$ to $10^6 s^{-1}$) were such that multiple cracks were activated simultaneously and not a single or few large cracks, as one might observe under low rate conditions. Lower strain rate would increase the size of the largest cracks and the current simulation box will not be able to account for it. The inability to activate multiple defects might lead to larger fragments at low rates, in which case the size of the fragments might be larger than the continuum scale and the macroscopic conditions might dominate the problem. Low rate fragmentation is also not the focus of the current work.\par
The transition wing crack length appears to be more sensitive to defect spacing than it is to defect size (Eq: \ref{eq:Trans_Dmg_Fit_Ran}  \& \ref{eq:PhenTrans90}). This might suggest material modification in favour of controlling the defect spacing rather than defect size to obtain desirable behaviour. Future work should focus on microstructural dependence of fragmentation, granular transition, flow and subsequent overall material behaviour under high rate conditions.\par
Most of the simulations in the study correspond to a randomly oriented defect distribution. However for the case of a fixed defect orientation along the most favourable direction, a transition equation has been similarly predicted (Eq: \ref{eq:Trans_Dmg_Fit_Del}), and it has a similar form to that of the random defect orientation case (Eq: \ref{eq:Trans_Dmg_Fit_Ran}). A more general way of explaining granular transition is in terms of active defects. For sliding crack models, the activation of defects depends upon the defect orientation, defect size, the stress field, the strain rate, the crack face friction coefficient  and the fracture toughness. While comparing between different initial defect orientations, the same number of defects which are able to overcome the crack face frictional constraint should have similar transitional response. In the absence of confinement, cracks with orientations greater than the friction angle can be activated. It is worth noting that overcoming the crack face friction ($\mu_{flaw}$) is not a sufficient condition for wing crack growth. However, we will denote these cracks as active cracks henceforth.
So, the density of active defects is $\eta_{active}=\frac{tan^{-1}(\mu_{flaw})}{\pi/2}\eta$.
From Eq: \ref{eq:Damage_2D_Dis} \& \ref{eq:Final_l} and for uniform crack size distribution, 
\begin{equation} 
  \label{eq:Active_Total_Crack_Rel}
l_f=\sqrt{E[l_w^2]} \implies l_f=\sqrt{E[{{l_w}_{active}}^2]\frac{\eta_{active}}{\eta}} \implies l_f={l_f}_{active}\sqrt{\frac{tan^{-1}(\mu_{flaw})}{\pi/2}},
\end{equation}
where, ${l_w}_{active}$ and ${l_f}_{active}$ are the $l_w$ and $l_f$ equivalent for active defects. Also, it can be assumed that the initial defect size is independent of the orientation distribution. 
Using these relations, Eq: \ref{eq:Trans_Dmg_Fit_Ran} can be rewritten as:
\begin{equation} 
  \label{eq:Trans_Dmg_Fit_Ran_Active}
{l_f}_{active}=1.354\bigg(\frac{\pi/2}{tan^{-1}(\mu_{flaw})}\bigg)^{1/6}\eta_{active}^{-1/3}-0.6977\bigg(\frac{\pi/2}{tan^{-1}(\mu_{flaw})}\bigg)^{1/2}l_i
\end{equation}
For $\mu_{flaw}=0.8$, Eq: \ref{eq:Trans_Dmg_Fit_Ran_Active} can be written as ${l_f}_{active}=1.5588\eta_{active}^{-1/3}-1.0645l_i$. \par
This is very similar to Eq: \ref{eq:Trans_Dmg_Fit_Del} in which all the similar sized defects have the same orientation and are activated simultaneously. In reality, the exact nature of the transition equation will also depend on the correlation between micro-structural defect density, size and orientation.

\section{Conclusion}

A physically based granular transition criterion for continuum models of high rate impact of brittle ceramics has been proposed. The model assumes near-instantaneous granular transition and suggests that a certain combination of state variables need to meet a certain threshold for fragmentation and transition to a granular state. This transition criterion serves as a switch in continuum codes for brittle dynamic fragmentation that activates granular physics. The outputs of the model also help characterize the initial conditions for granular mechanics as a function of initial defect characteristics. A simple phenomenological transition model also proposes a similar form of transition equation without delving into the mechanics of crack growth. 

\section{Acknowledgement}
Research was sponsored by the Army Research Laboratory and was accomplished under Cooperative Agreement Number W911NF-12-2-0022. The views and conclusions contained in this document are those of the authors and should not be interpreted as representing the official policies, either expressed or implied, of the Army Research Laboratory or the U.S. Government. The U.S. Government is authorized to reproduce and distribute reprints for Government purposes notwithstanding any copyright notation herein.
\par
The simulations were performed using the high performance computing cluster at the \href{https://www.marcc.jhu.edu}{Maryland Advanced Research Computing Center(MARCC)}.\par
The authors would like to thank Prof. K.T. Ramesh, Prof. Mark Robbins and everyone else involved in the \href{https://hemi.jhu.edu/cmede/}{CMEDE} Ceramics Modelling group for their valuable contributions through discussions and suggestions.

\bibliographystyle{apacite}
\bibliography{References.bib}
\end{document}